%% file: main.tex
\documentclass[sigconf,acmsmall]{acmart}

\AtBeginDocument{%
  }

\usepackage{graphicx}
\usepackage{algorithm}
\usepackage{algpseudocode}
\usepackage{listings}
\usepackage{fancyvrb}
\usepackage{tikz}
\usepackage{tcolorbox}
\usetikzlibrary{shapes, arrows, fit, calc, arrows.meta, automata}
\usepackage{multirow}
\usepackage{booktabs}
\usepackage{amsmath}
\usepackage{tcolorbox}
\usepackage{float}
\tcbuselibrary{skins}
\usepackage{prettyref}
\usepackage{tabularx}
\usepackage{adjustbox}

\newrefformat{eq}{Eq.~(\ref{#1})}
\newrefformat{sec}{Section~\ref{#1}}
\newrefformat{ex}{Example~\ref{#1}}
\newrefformat{tab}{Table~\ref{#1}}
\newrefformat{alg}{Algorithm~\ref{#1}}
\usepackage{subcaption}

\usepackage[table]{xcolor}

\newif\ifcommentson
\commentsontrue
\newcommand{\julian}[1]{\ifcommentson\textcolor{blue}{Julian: #1}\fi}

\usepackage{xspace}
\newcommand{\ourtool}{\textsc{VerIbmc}\xspace}

\newcolumntype{C}[1]{>{\Centering\arraybackslash}p{#1}}

\setcopyright{none}
 
\begin{document}

\title{Neuro-Symbolic Software Verification: Hyper-charging Local Language Models with Symbolic Reasoning at Scale}


\author{Muhammad A. A. Pirzada}
\email{muhammad.pirzada@postgrad.manchester.ac.uk}
\affiliation{
  \institution{The University of Manchester}
  \city{Manchester}
  \country{UK}
}
\orcid{0009-0005-2440-7547} 

\author{Julian Parsert}
\email{julian.parsert@gmail.com}
\affiliation{
  \institution{RPTU Kaiserslautern}
  \city{}
  \country{Germany}
}
\orcid{0000-0002-5113-0767} 

\author{Weiqi Wang}
\email{Weiqi.Wang-2@postgrad.manchester.ac.uk}
\affiliation{%
  \institution{The University of Manchester}
  \city{Manchester}
  \country{UK}
}
\orcid{0009-0005-8247-5994}

\author{Konstantin Korovin}
\email{Konstantin.Korovin@manchester.ac.uk}
\affiliation{%
  \institution{The University of Manchester}
  \city{Manchester}
  \country{UK}
}
\orcid{0000-0002-0740-621X}

\author{Lucas C. Cordeiro}
\email{lucas.cordeiro@manchester.ac.uk}
\affiliation{
  \institution{The University of Manchester}
  \city{Manchester}
  \country{UK}
}
\orcid{0000-0002-6235-4272} 

\renewcommand{\shortauthors}{Pirzada, Parsert, Wang, Korovin, Cordeiro}

\begin{abstract}
Loop invariant synthesis remains a central and pivotal bottleneck in formal software verification. Recent LLM-based Neuro-Symbolic tools have achieved impressive solve rates. However, these tools rely on proprietary, often expensive cloud APIs, which constitute a hurdle for privacy-sensitive industrial deployments where the source code cannot leave the organisation or where cost is a factor. We present \ourtool, a neuro-symbolic pipeline that pairs symbolic invariant generation with locally deployable open-weight language models with the \textsc{ESBMC} verification tool. Our pipeline combines a deterministic symbolic invariant synthesis phase with an iterative LLM refinement loop driven by structured verifier feedback. In addition, we provide two types of pipelines that differ in their prompting strategy: \textit{Chain-of-Thought} vs. \textit{Tree-of-Thought}.

We conduct an extensive experimental evaluation with five open-weight models (ranging from 7B to 120B parameters) across five benchmark families comprising of 520 problems  (499 after excluding 21 with unavoidable overflow). Overall, the best single configuration (GPT-OSS-120B) solves $431$ of $499$ problems ($86.4$\%). Additionally, on the four benchmark suites shared with the strongest cloud-API tools, \ourtool{} is competitive running only on a single local machine. The evaluation shows symbolic invariant synthesis solves $75$ problems without any LLM call and yields up to $+35$ additional problems for the weakest model. Importantly, all inference runs entirely on a single local machine using open-weight models -- no cloud API or proprietary model is required. Overall, we demonstrate that a neuro-symbolic approach based on LLMs can be used effectively for invariant synthesis in a privacy-preserving and energy-efficient manner, without having to resort to expensive proprietary frontier models locked behind APIs.
\end{abstract}




\maketitle

\section{Introduction}
Software systems increasingly underpin safety-critical and security-sensitive domains, including but not limited to: transportation \cite{koopman2017autonomous, charette2009car,fraser2024crowdstrike}, healthcare \cite{leveson1993investigation}, and finance \cite{kirilenko2017flash}. In these settings, correctness is not merely desirable but essential. Formal verification provides mathematically rigorous guarantees that programs satisfy their specifications by reasoning about their behavior for all possible inputs. Deductive verification, in particular, has become a key approach for reasoning about real-world software. In this work, we target the problem of loop invariant synthesis, as this is a central bottleneck for verification. Notably, our techniques are not language specific. However, its effectiveness depends crucially on the availability of appropriate tooling for loop invariant verification~\cite{DBLP:journals/csur/FuriaMV14}.

Despite decades of progress in program analysis and not least due to its undecidability, automatic loop invariant synthesis remains a central bottleneck in formal verification. Symbolic techniques such as abstract interpretation~\cite{DBLP:conf/popl/CousotH78}, constraint solving~\cite{DBLP:conf/vmcai/BeyerHMR07,DBLP:conf/cav/ColonSS03,DBLP:conf/vmcai/HumenbergerJK18}, and other methods provide partial solutions but often struggle with scalability, expressiveness, or the diversity of real-world code. More recently, data-driven invariant synthesis based on ``guess-and-check'' methods have become popular~\cite{DBLP:conf/popl/0001NMR16,DBLP:journals/fmsd/SharmaA16,DBLP:conf/esop/0001GHALN13,DBLP:conf/sigsoft/GiacobbeKP22,DBLP:conf/sigsoft/Xu0W20,DBLP:conf/pldi/YaoRWJG20,DBLP:conf/issta/Yu0023}.

Recently, large language models (LLMs) have significantly impacted software engineering practice~\cite{DBLP:journals/tosem/HouZLYWLLLGW24}. They are now widely used for tasks such as code generation, summarization, debugging, and test creation~\cite{DBLP:conf/nips/YangJWLYNP24,DBLP:conf/iclr/JimenezYWYPPN24}. Their ability to learn statistical and semantic regularities from large code corpora has also made them attractive for formal methods research. In particular, LLMs have shown promise in generating candidate loop invariants and other verification artifacts by leveraging learned patterns of program structure and behavior~\cite{DBLP:conf/emnlp/ChakrabortyLFLM23,DBLP:journals/corr/abs-2311-07948,wu2024lemur,DBLP:conf/kbse/WuC0W0M24}. This has opened a new direction in which LLM-based approaches complement traditional symbolic reasoning.
However, many existing approaches rely on proprietary, closed models accessed via external APIs. This introduces several limitations in the context of formal verification of production software. First, source code must often be transmitted to third-party services, raising serious confidentiality and compliance concerns~\cite{DBLP:journals/corr/abs-2512-10169,DBLP:conf/iclr/Mireshghallah0Z24,DBLP:journals/csur/DasAW25}. Second, reliance on external providers reduces reproducibility and complicates integration into controlled verification pipelines. Third, large-scale models are computationally and financially expensive and are associated with significant energy consumption at both training and inference, thereby raising sustainability concerns and limiting their practicality for continuous or large-scale verification workflows~\cite{DBLP:conf/green/DingS24,DBLP:conf/hotcarbon/ChienLNRSW23}.

These challenges motivate the use of open-weight and open-source language models that can be deployed locally within an organization’s infrastructure. Such models enable full control over data, eliminate the need to expose proprietary code externally, and improve reproducibility of verification results. Importantly, smaller and medium-sized models also offer a significant advantage in energy efficiency, making them better suited for sustained use in industrial software verification pipelines, where repeated analysis and iteration are required.

In this work, we investigate the use of locally deployable, open-weight language models for loop invariant synthesis in C programs. The pipeline is language-agnostic by design: it interacts with the verifier only through a loop-invariant-checking interface, and \textsc{ESBMC} supports languages beyond C/C++ (e.g. Python) through dedicated front-ends. Our evaluation focuses on C for comparability with existing tools; we do not claim empirical results on other languages.
Our goal is to support formal verification of production software in environments where privacy, reproducibility, and resource efficiency are critical requirements. By operating entirely on internal infrastructure, our approach avoids exposing external data while enabling seamless integration into existing deductive verification toolchains. We present a complete pipeline that integrates locally served LLMs into a loop invariant synthesis and verification loop. The system generates candidate invariants, validates them with a formal verification backend, and iteratively refines them in response to counterexamples or proof failures. When the iteration budget is exhausted without finding a valid invariant, the pipeline then reports the outcome as unknown and terminates. The method is always sound but incomplete due to the inherent undecidability of the problem. This structured interaction allows the verifier to guide the model, improving both precision and robustness of generated invariants.
We evaluate our approach on a representative set of verification benchmarks and demonstrate that locally deployed open-weight models can effectively support loop invariant generation across a wide range of programs. Our results show that when properly integrated into a formal verification loop, such models can significantly reduce annotation effort while maintaining strong performance on verification tasks.

Overall, this work demonstrates that combining LLMs with symbolic reasoning can be applied to invariant synthesis with great effect. When using relatively small open-weight models as part of this neuro-symbolic approach, we can verify programs in a privacy-preserving, energy-efficient manner, without relying on proprietary cloud-based systems. With these verification pipelines we take a step toward scalable, reproducible, and industry-ready AI-assisted formal verification of production software.
To summarize, our contributions are:

\begin{itemize}
    \item We present a neuro-symbolic method that integrates LLMs with symbolic reasoning using ESBMC as a verification oracle for automated loop invariant synthesis in C programs.
    \item We design and experiment with inference strategies spanning pure LLM, symbolic-guided, and Tree-of-Thoughts (ToT) variants, providing a systematic comparison of neural and neuro-symbolic approaches.
    \item We demonstrate that symbolic feedback (provable invariant atoms fed as context to the LLM) provides consistent gains for weaker models, yielding up to 35 additional solved benchmarks over a pure-LLM baseline (Llama-3.1-8B: 307$\to$342). 
    \item We describe and evaluate the effect of different prompting strategies including, Chain-of-Thought (CoT) and ToT prompting on loop invariant synthesis. We also experiment different reasoning styles in the prompting stages (Inductive, Hoare Logic, Horn Clause, and direct derivation).
    

    \item We conduct a large-scale evaluation across five language models, four inference strategies, and five benchmark families (520 problems), including SV-COMP benchmarks, providing a comprehensive assessment of LLM-guided invariant synthesis at scale (10{,}400 per-problem outcomes). 
\end{itemize}

\section{Theoretical Background}
In deductive verification, the correctness problem is reduced to proving logical formulas called verification conditions (VCs). These conditions encode the semantics of the program together with its specification. If all generated verification conditions are valid, then the program is considered correct with respect to the specification. Modern verification frameworks usually translate programs into an intermediate logical representation based on first-order logic, often extended with theories such as arithmetic, arrays, bit-vectors, or algebraic data types. Automated reasoning about these formulas is delegated to Satisfiability Modulo Theories (SMT) solvers. Popular SMT solvers such as Z3 \cite{de2008z3}, Boolector \cite{brummayer2009boolector}, MathSAT \cite{cimatti2013mathsat5}, Yices \cite{dutertre2006yices}, and Bitwuzla \cite{niemetz2023bitwuzla} are widely used in program verification.
In the context of deductive verification, the verifier generates formulas representing correctness obligations and submits them to an SMT solver. If the solver proves the formulas unsatisfiable under the negation of the specification, then the corresponding verification condition holds. Additionally, SMT solvers provide efficient reasoning procedures for many theories relevant to software verification and therefore form the backbone of many modern verification tools.

Furthermore, model checking~\cite{DBLP:reference/mc/2018} is another major verification paradigm. Instead of proving correctness via symbolic logical derivations alone, model checking systematically explores the state space of a system to determine whether a specification holds.
Bounded Model Checking (BMC) is a formal verification technique introduced by Biere et al.~\cite{biere1999symbolic,DBLP:series/faia/Biere21} that systematically checks whether a system violates a given safety property within a finite number of execution steps, known as the \emph{bound} $k$. The fundamental idea is to unroll the program's transition relation $k$ times, producing a propositional formula that is satisfiable if and only if an error state is reachable within $k$ steps. 
A central limitation of plain BMC is its inherent incompleteness: a clean result for bound $k$ does not guarantee correctness for $k+1$ or beyond. To address this, \emph{$k$-induction}~\cite{sheeran2000checking} extends BMC with a proof step: if no violation exists within $k$ steps \emph{and} every $k$-step execution prefix that satisfies the invariant candidate also satisfies it at step $k+1$, the property is proven for all depths. However, $k$-induction only succeeds when the candidate invariant is already strong enough; for programs with complex loop behaviour, a strong \emph{loop invariant} must be supplied to strengthen the inductive hypothesis. This is precisely where invariant inference becomes critical: a sufficiently strong invariant, when assumed at the loop head, allows the BMC engine to complete a full inductive proof without requiring deeper unrolling.

\subsection{Loop Invariants}
First of all, a \emph{loop} is a control-flow construct that repeatedly executes a $\mathit{body}$, as long as a guard $g$ evaluates to true~\cite{hoare1969axiomatic}. Since the number of iterations is either large or unbounded, reasoning about a loop requires a finite characterization of its behavior -- that holds across all iterations. This is where loop invariants are extremely beneficial. A \emph{loop invariant} is a predicate $\varphi$ that must satisfy three
conditions~\cite{hoare1969axiomatic,floyd1993assigning}:
\begin{enumerate}
    \item \textit{Initiation}: $\varphi$ holds before the first iteration
          ($\mathit{pre} \Rightarrow \varphi$).
    \item \textit{Consecution}: $\varphi$ is preserved by every iteration
          ($\{\varphi \wedge g\}\; \mathit{body}\; \{\varphi\}$).
    \item \textit{Sufficiency}: $\varphi$ together with the negated loop guard
          implies the required post-condition
          ($\varphi \wedge \neg g \Rightarrow \psi$).
\end{enumerate}

A loop invariant satisfying all three conditions is called an \emph{inductive loop invariant}. A predicate that merely holds at the loop entry point but is not preserved by the loop body is non-inductive and therefore insufficient to discharge the associated verification condition.

\paragraph{The Challenge of Invariant Synthesis}
Invariant synthesis is undecidable in general~\cite{rice1953classes,floyd1993assigning,hoare1969axiomatic}: no algorithm can, given an arbitrary program and post-condition, always return an adequate loop invariant. The difficulty is not confined to pathological programs. Even for loops over integer variables with simple control flow, an adequate invariant may require non-linear arithmetic, conjunctions of many linear constraints, or ghost variables absent from the original source. Hence, practical invariant synthesis tools must navigate a search space whose membership problem is itself undecidable, trading completeness for tractability on the program classes that arise in practice.

\paragraph{ESBMC} (Efficient SMT-Based Context-Bounded Model Checker)~\cite{gadelha2019esbmc, cordeiro2011smt, Sa_Menezes_ESBMC_7_4_Harnessing} is an industrial-strength open-source verification tool for C/C++ programs. It front-ends the LLVM/Clang compiler infrastructure, performs pointer analysis, arithmetic overflow and array bounds checking, and encodes the resulting verification conditions in SMT-LIB2, discharging them via one of several back-end solvers (Z3~\cite{de2008z3}, Boolector~\cite{brummayer2009boolector}, MathSAT~\cite{cimatti2013mathsat5}, Yices~\cite{dutertre2006yices}, Bitwuzla~\cite{niemetz2023bitwuzla}). Beyond plain BMC, ESBMC supports $k$-induction and a loop-invariant-checking mode that compiles the program to ESBMC's GOTO representation and injects a user-supplied predicate at the loop head, establishing correctness for all loop iterations when a sufficient invariant is supplied. Throughout this paper, ESBMC serves as the formal verification back-end for every invariant candidate produced by the LLM-based components of our approach. We invoke ESBMC exclusively in loop-invariant-checking mode with $k$-induction as the underlying proof rule; we do not rely on plain BMC, and our soundness argument therefore does not depend on a bounded unrolling.

\subsection{LLMs}

LLMs are deep neural sequence models, trained on web-scale corpora of natural language and source code via a \emph{next-token prediction} objective. Popular network types are transformer models, which allow each token in a sequence to attend to every other token, enabling the model to capture long-range syntactic and semantic dependencies that earlier recurrent architectures struggled to represent. Scaling the number of parameters, training tokens, and compute following the empirical laws described by Kaplan et al.~\cite{kaplan2020scaling} and subsequently refined as \emph{chinchilla scaling}~\cite{hoffmann2022training} has produced a family of increasingly capable models: GPT ~\cite{achiam2023gpt}, Claude ~\cite{anthropic2024claude}, Gemini~\cite{team2023gemini}, Llama ~\cite{touvron2023llama, grattafiori2024llama}, and Alibaba's Qwen \cite{bai2023qwen, qwen2025qwen25} and DeepSeek-Coder~\cite{guo2024deepseek}, among others.

Furthermore, a defining capability of sufficiently large models is \emph{in-context learning} (ICL)~\cite{brown2020language}: by conditioning the model on a prompt that contains a handful of input-output examples, it generalises to new inputs in the same format without any gradient update to its parameters. This is distinct from traditional supervised learning and is particularly valuable in formal verification, where labelled (program, invariant) pairs are scarce and domain-specific fine-tuning datasets are expensive to construct. 
For code-related tasks specifically, LLMs trained on large repositories of open-source software have internalised the syntactic conventions, common idioms, and proof patterns of widely used programming languages and verification frameworks. Studies such as HumanEval~\cite{chen2021evaluating} and SWE-bench~\cite{DBLP:conf/iclr/JimenezYWYPPN24} demonstrate that frontier models can solve a significant fraction of competitive programming and real-world bug-fixing tasks. In the context of loop invariant inference, this prior exposure to annotated code, SMT-LIB formulae, and formal specification languages means that LLMs can often produce plausible invariant candidates from program text alone, without requiring domain-specific training.

Despite these strengths, LLMs are fundamentally probabilistic predictors and are known to \emph{hallucinate}: they produce plausible-sounding but factually or logically incorrect outputs with non-trivial probability, including syntactically valid but semantically incorrect program assertions. In a verification context, a hallucinated invariant that is not inductively valid would cause a sound verifier to reject it, but could also mislead a developer who does not independently check it. This makes LLMs unsuitable as standalone verifiers. The dominant design pattern in the literature is therefore a \emph{guess-and-check} architecture: the LLM acts as a heuristic proposal function that generates candidate invariants, while a formal tool verifies the given candidate. 
\subsection{Prompt Engineering: CoT and ToT Reasoning}
Prompting describes the presentation of a task description which is given as an input to the LLM. Standard prompting often fails on tasks that require multi-step reasoning, because the model is forced to compress all intermediate reasoning into a single generation step. Two prompting strategies have emerged as principled solutions to this limitation, both of which are directly applicable to loop invariant inference: CoT and ToT prompting.

\paragraph{CoT prompting~\cite{wei2022chain}} This elicits explicit intermediate reasoning by including, in the few-shot prompt, example solutions that demonstrate step-by-step derivation rather than simply showing input-output pairs. The model is implicitly encouraged to emit a sequence of reasoning steps (i.e. a \emph{chain of thought}) before producing its final answer. Wei et al.~\cite{wei2022chain} showed that CoT dramatically improves performance on arithmetic, commonsense, and symbolic reasoning benchmarks, and that the benefit is an emergent property of models with sufficiently many parameters (approximately $100$B in the original study, though subsequent work has replicated the effect in smaller models with better training). In the invariant synthesis context, a CoT prompt might instruct the model to: (i) identify the loop's modified variables and their ranges; (ii) determine how each variable evolves across iterations; (iii) derive a relational constraint that is preserved by the loop body; and (iv) express that constraint as a formal assertion. This structured decomposition mirrors the manual reasoning process of a verification engineer and has been shown to improve the syntactic validity and semantic correctness of generated invariants compared to direct prompting~\cite{wu2024lemur, DBLP:conf/kbse/WuC0W0M24, DBLP:conf/cav/WenCSXQHLCT24}.

A particularly effective variant is \emph{zero-shot CoT}~\cite{kojima2022large}, in which the model is prompted with the phrase ``Let's think step by step'' (or a domain-appropriate equivalent such as ``Let's derive the loop invariant step by step'') without providing explicit few-shot examples. This eliminates the need to curate high-quality example solutions and has been shown to achieve performance close to few-shot CoT on many reasoning benchmarks, making it practical in settings where annotated (program, invariant) examples are scarce.

\emph{ToT} prompting~\cite{yao2023tree} generalises CoT via maintaining a \emph{tree} of partial reasoning states rather than a single linear chain. At each node in the tree, the model generates $k$ alternative continuations each representing a different hypothesis about the next step of the derivation. A \emph{state evaluator} (which may be the LLM itself, a symbolic checker, or a learned heuristic) scores each continuation and determines which branches to expand further, implementing a best-first or breadth-first search over the space of reasoning paths. ToT was introduced by Yao et al.~\cite{yao2023tree} and demonstrated significant improvements over CoT on tasks requiring planning and backtracking, such as the Game of 24 and creative writing with structural constraints.


\section{The \ourtool{} System}
\label{sec:system}
\label{sec:methodology}

\label{sec:overview}


\begin{figure}
    \centering
    \includegraphics[width=\columnwidth]{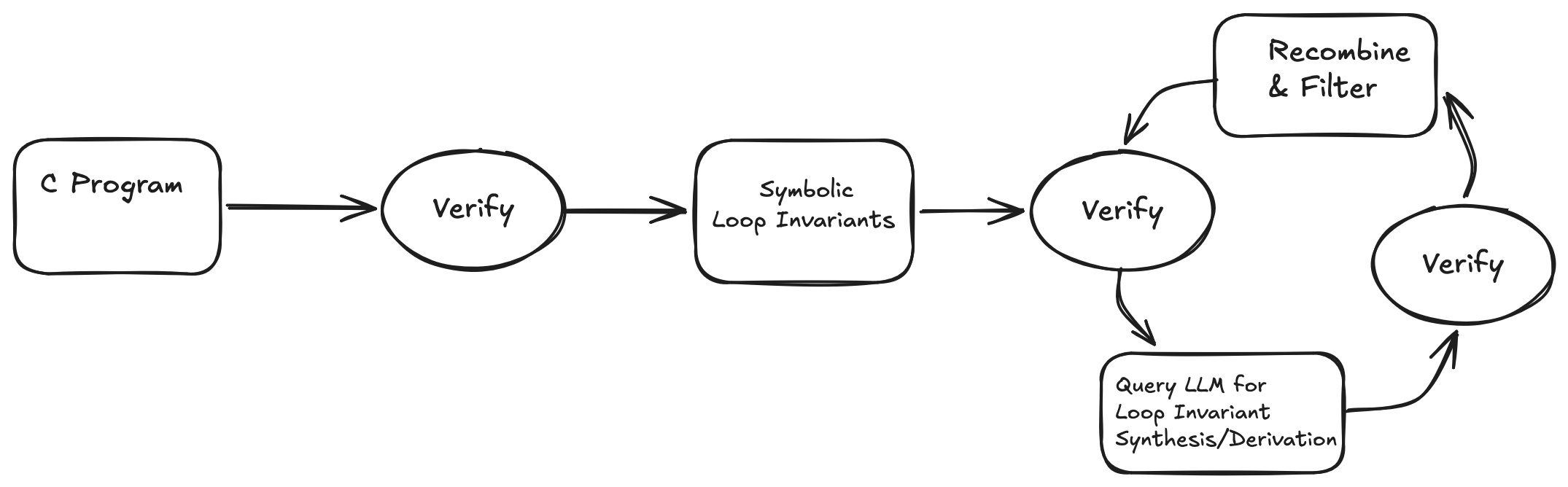}
    \caption{\ourtool Flow Diagram}
    \label{fig:verIbmc}
\end{figure}

\input{content/system}


\input{content/experiments}

\section{Threats to Validity}
\label{sec:threats}
Like the state-of-the-art tools we compare against, our pipeline reasons about programs under a real-arithmetic (LRA/LIA-over-$\mathbb{Z}$) semantics rather than over C's fixed-width machine integers. This makes the head-to-head comparison fair -- every system in the evaluation is solving the same abstract problem on the same benchmarks. For this comparison, we therefore operate under the same abstract semantics as the competing tools, which means overflow bugs are not checked in this evaluation. This is a deliberate choice for comparability, not a fundamental limitation: ESBMC natively supports bit-precise reasoning via SMT bitvector theory, so \ourtool{} can detect arithmetic overflows when configured to do so -- a capability that the cloud-API tools we compare against do not offer.

\paragraph{Model bias.}
GPT-OSS-120B and GPT-OSS-20B belong to the same model family~\cite{agarwal2025gpt} and share architectural design choices and training provenance; their results are therefore not fully independent.
The scaling trend we observe -- larger model yields higher solve rate -- may partly reflect shared inductive biases rather than a universal property of open-weight models in general.
All five models were served through Ollama at default generation parameters; we did not tune temperature or sampling settings per model or per benchmark family.
Decoding configurations that are suboptimal for a given model may understate its performance, and the relative rankings we report could shift under different hyperparameter settings.
Additionally, all five benchmark families used in this evaluation are publicly available -- C2I, L4I-SVC, L4I-SY, and NL are distributed alongside their respective tools, and SVC problems are drawn from the public SV-COMP archive.
Since all five models are trained on internet-scale corpora, it is likely that some programs, their invariants, or the papers introducing them have appeared in pre-training data.
This could artificially inflate per-problem LLM performance on those families, and we did not attempt to detect or quantify contamination.

\paragraph{Non-determinism and experimental variance.}
LLM generation is stochastic.
Each (model, strategy, problem) triple in the full four-strategy grid (Tables~\ref{tab:results_veribmc_cot}--\ref{tab:results_veribmc_tot}) was evaluated in a \emph{single} run; this applies to LLM-Only, Basic, LLM-Only-ToT, and Basic-ToT alike.
To quantify sampling sensitivity for the \emph{best-performing} strategy, we replicated the Basic strategy specifically: in addition to the original run, we conducted two further independent full runs of Basic across all five models and all eleven benchmark families (2{,}600 model--problem pairs per run), giving three runs of Basic in total.
We report the best result across the three runs for each model as the main Basic result in Table~\ref{tab:results_veribmc_cot}; for GPT-OSS-120B this corresponds to Run~1 (431 problems), while all other models achieve their best in Run~0.
Table~\ref{tab:reproducibility} summarises all three runs.
Grand totals across the three runs are 1{,}928, 1{,}913, and 1{,}918; per-model standard deviations range from $\sigma = 0.6$ (GPT-OSS-20B) to $\sigma = 5.3$ (Llama-3.1-8B, the weakest model and therefore most sensitive to sampling fluctuation).
The between-run range for the strongest model, GPT-OSS-120B, is only 3 problems (428--431), well within the 89-problem gap between GPT-OSS-120B and GPT-OSS-20B reported in Table~\ref{tab:results_veribmc_cot}.
All three grand totals fall within a 15-problem window (1{,}913--1{,}928) against a total of 2{,}600, confirming that the rank ordering of models and the strategy-level findings are robust to LLM non-determinism.
Problems at the boundary of model capability may still produce different outcomes under different random seeds; differences of fewer than ten problems between \emph{strategies} should be treated as indicative rather than definitive.
The 600\,s wall-clock timeout introduces a further source of variance: host machine load during a run can shift which problems are decided within budget.

\begin{table}[t]
\centering
\begin{tabular}{lrrrrc}
\toprule
Model & Run 0 & Run 1 & Run 2 & Mean & $\sigma$ \\
\midrule
GPT-OSS-120B         & 428 & 431 & 428 & 429.0 & 1.7 \\
GPT-OSS-20B          & 424 & 423 & 424 & 423.7 & 0.6 \\
Qwen2.5-32B-Instruct & 382 & 380 & 381 & 381.0 & 1.0 \\
Qwen2.5-7B-Instruct  & 352 & 347 & 351 & 350.0 & 2.6 \\
Llama-3.1-8B         & 342 & 332 & 334 & 336.0 & 5.3 \\
\midrule
Grand Total          & 1928 & 1913 & 1918 & 1919.7 & 7.6 \\
\bottomrule
\end{tabular}
\caption{Solve counts for the Basic strategy across three independent runs (2{,}600 model--problem pairs per run; 520 problems $\times$ 5 models). $\sigma$ is the sample standard deviation across the three runs.}
\label{tab:reproducibility}
\end{table}

\paragraph{Benchmark bias.}
Three of the four shared benchmark families were curated by or introduced alongside the tools we compare against: \texttt{ibmc\_code2inv} derives from \textsc{Code2Inv}~\cite{si2018learning}, \texttt{L4I-SVC} and \texttt{L4I-SY} were introduced alongside \textsc{LaM4Inv}~\cite{DBLP:conf/kbse/WuC0W0M24}, and \texttt{clause2inv\_NL} was introduced alongside \textsc{Clause2Inv}~\cite{DBLP:journals/pacmse/CaoWXYWCM25}.
Benchmarks curated by competing tools may implicitly reflect their representational strengths, potentially giving those tools a distributional advantage on their respective sets. Furthermore, these benchmarks were, in part, syntactically restricted to conform to the input format of the tested tools. To mitigate this bias, we also considered a wider range of benchmarks provided by the SV-COMP that are collected from various sources. Furthermore, these benchmarks do not fall under the same syntactic restrictions.

\paragraph{Benchmark representativeness.}
The majority of benchmark problems in our evaluation are single-loop programs over scalar integer variables. Notably, the SV-COMP subset includes programs with multiple loops, arrays, and pointer arithmetic (see \prettyref{sec:rq2}).
This design choice isolates arithmetic invariant reasoning but limits generalisability to industrial software.
Programs with heap-allocated data structures, pointer aliasing, concurrency, floating-point arithmetic, or complex multi-procedure control flow lie outside the current scope.
Nevertheless, the SV-COMP subset substantially relaxes the single-loop and scalar integer restrictions. As a result \ourtool{} handles arrays, nested loops, and function calls. This is enabled through the use of  \textsc{ESBMC}'s advanced C frontend that supports a wide range of C features. In particular, of the 154 SV-COMP programs, 36 contain multiple or nested loops that \ourtool{} can support but that are inexpressible for some competing tools (see \prettyref{tab:svc-limitation}). As a result, the SV-COMP problems that fall outside the expressible fragment of some SOTA tools have no full comparison.

\paragraph{Cross-tool comparison on SV-COMP}
For the SV-COMP set, \textsc{Clause2Inv} and \textsc{LaM4Inv} were run by us using their required external paid APIs, because their published evaluations did not cover this benchmark.
These runs were conducted at different times, used the API model versions available at the time of our evaluation, and may have executed on different backend infrastructure from the runs reported in their respective papers. LLM API outputs are non-deterministic and API model versions change silently over time; these cross-tool figures should therefore be treated as indicative.
The comparable-subset comparison (Table~\ref{tab:results_sota}) uses the same 43 programs under identical evaluation conditions for all three invariant-driven methods and is the most controlled tier of the SV-COMP comparison. 

\paragraph{ToT scoring function.}
The ToT strategy-scoring function (Eq.~\ref{eq:score}) assigns fixed weights of $0.6$, $-0.3$, and $0.1$ to provable, disprovable, and unknown atoms respectively.
These weights were chosen heuristically and were not subjected to ablation.
The top-2 strategy selection in Stage~1, and the re-ranking every six rounds in Stage~2, both depend directly on this scoring function.
A different weighting could change which two strategies are promoted to the refinement loop for a given program, potentially altering which problems ToT solves or fails on and shifting the diversity--accuracy trade-off boundary reported in \prettyref{sec:rq3}.
The reported ToT results should therefore be understood as specific to this particular scoring configuration rather than as a property of the ToT architecture in general.

\section{Related Work}
\paragraph*{Traditional Methods.}

\emph{Abstract interpretation}~\cite{cousot1977abstract} computes sound over-approximations over abstract domains (intervals, octagons, polyhedra~\cite{DBLP:conf/popl/CousotH78}), but precision is domain-limited. \emph{Constraint-based} synthesis~\cite{DBLP:conf/cav/ColonSS03,sankaranarayanan2004non} encodes initiation and consecution as constraints over a fixed template, requiring the user to anticipate the invariant's form. \emph{Houdini}~\cite{flanagan2001houdini} computes the largest simultaneously-inductive subset of a candidate set but cannot synthesise new predicates. \emph{IC3/PDR}~\cite{bradley2011sat,een2011efficient} incrementally strengthens reachability over-approximations using counterexamples to inductiveness; it is complete for finite-state systems and extends to infinite-state via SMT~\cite{hoder2012generalized}.

\paragraph*{Neuro-Symbolic Methods.}
Neuro-Symbolic approaches pair a neural proposal component (GNN or LLM) with a symbolic back-end that checks soundness and returns structured feedback (counterexamples, unsat cores). This \emph{generate-and-check} architecture is sound by construction: any accepted invariant is logically guaranteed, regardless of how it was generated. Earlier RL-based work (\textsc{Code2Inv}~\cite{si2018learning}) required thousands of solver queries during training and suffered from sparse binary rewards; LLM-based approaches leverage pretrained knowledge to reduce oracle queries substantially. All five evaluated systems fall within this paradigm, differing in neural architecture, feedback format, and use of fine-tuning.

\textsc{Code2Inv}~\cite{si2018learning} encodes programs as graphs (data-flow, control-flow, syntactic containment), applies a GNN to produce invariant candidates, and updates via policy-gradient RL using Z3 outcomes as rewards. It solved 106/133 benchmark programs, outperforming template-based methods, and established the de facto evaluation suite used by all subsequent work. The tool is limited to linear arithmetic invariants over scalar integers.
IC3Syn~\cite{cao2026synthesizing} combines an IC3/PDR frame-based controller with LLM-generated blocking clauses to synthesise inductive invariants for distributed protocols specified in TLA+; unlike \ourtool{}, it targets protocol-level transition systems rather than C loop programs and relies on cloud-hosted frontier models via external APIs.

\textsc{LEMUR}~\cite{wu2024lemur} is the first framework to provide a formal calculus for integrating LLMs into automated program verification. Both GPT-3.5-turbo and GPT-4 are supported as oracle backends, with GPT-4 demonstrating superior performance; the program and any solver counterexample are included in the prompt, and on failure, a \emph{repair} step applies backtracking to refine candidates before re-checking. LEMUR achieved state-of-the-art results on \textsc{Code2Inv} and non-linear benchmarks with no task-specific fine-tuning.

\textsc{ESBMC-ibmc}~\cite{10.1145/3691620.3695512} integrates LLM-generated invariants into the \textsc{ESBMC} BMC pipeline by modifying the control flow graph (CFG) rather than unrolling loops. A portion of the CFG representing a loop is replaced by a node asserting LLM-generated invariants, which are validated by \textsc{Vampire}~\cite{kovacs2013first}, a first-order theorem prover, to ensure soundness. This transforms programs into loop-free variants that \textsc{ESBMC} can then verify. \textsc{VerIbmc} eliminates this external dependency: all invariant checking is discharged directly through \textsc{ESBMC}'s loop-invariant-checking mode with $k$-induction, requiring no additional theorem prover.



\textsc{LaM4Inv}~\cite{DBLP:conf/kbse/WuC0W0M24} integrates LLMs with BMC in a closed-loop \emph{query-filter-reassemble} strategy. The LLM generates candidate invariants, which are filtered and fed into the \textsc{ESBMC} BMC pipeline; counterexamples from failed checks are returned to the LLM to guide refinement. \textsc{LaM4Inv} outperforms both traditional and prior LLM-based methods; a detailed quantitative comparison with \textsc{VerIbmc} is given in Section~\ref{sec:results}.

\textsc{Clause2Inv}~\cite{DBLP:journals/pacmse/CaoWXYWCM25} refines the standard guess-and-check framework into a \emph{generate-combine-check} framework. An LLM-based clause generator produces atomic clauses (simple expressions without logical connectives), which are stored and systematically combined by a counterexample-driven combinator. Each candidate conjunction is verified by an SMT solver; failures trigger further combination guided by counterexamples. This factorised architecture addresses the observation that existing approaches struggle with complete invariants due to the complexity of logical connectives, yet the individual clauses of the correct invariant typically appear in earlier guesses.

\textsc{LORIS} ~\cite{10.1145/3806652} identify a fundamental limitation of counterexample-based feedback: informing an LLM that a concrete variable assignment violates its candidate invariant is a blunt signal, because LLM errors are typically detail mistakes within an otherwise-correct reasoning strategy rather than global misdirections. \textsc{LORIS} addresses this by prompting the LLM to produce a step-by-step natural language proof of its proposed invariant, translating each proof step into a first-order logic implication via a dedicated \emph{Formalizer} LLM, and checking each implication with Z3~\cite{de2008z3}. When a specific implication fails, the model receives targeted feedback identifying exactly which reasoning step is invalid -- for example, that the claim $j > m \Rightarrow j = m+1$ in step~3 does not hold -- rather than a raw counterexample. LORIS achieves strong accuracy on overlapping benchmarks, with its advantage over clause-based approaches growing with model capability; see Section~\ref{sec:results} for a detailed comparison with \textsc{VerIbmc}.

\textsc{VerIbmc} and LORIS address complementary deployment scenarios.
LORIS requires transmitting source code to an external API and degrades
substantially with weaker models. \textsc{VerIbmc} targets environments where this is infeasible: all inference runs on locally-deployed open-weight models, and Phase~1 symbolic augmentation provides a deterministic floor that particularly benefits weaker models. Phase~1 also resolves a non-trivial fraction of benchmarks without any LLM call, a category LORIS cannot address (see Section~\ref{sec:results}). Li et al.\ also demonstrate that local-error feedback is complementary to LaM4Inv-style symbolic conjunct accumulation~\cite{DBLP:conf/kbse/WuC0W0M24}, suggesting that integrating LORIS-style reasoning feedback into \textsc{VerIbmc}'s Phase~2 prompt is a natural direction for future work (Section~\ref{sec:future}).

\paragraph*{LLM-Based Specification Synthesis.}
The generate-and-check paradigm extends naturally from loop invariants to broader verification artifacts. \textsc{AutoSpec}~\cite{DBLP:conf/cav/WenCSXQHLCT24} augments LLM-based specification synthesis with static analysis and program verification, iteratively refining Frama-C annotations until a formal checker accepts them. Granberry et al.~\cite{DBLP:conf/ifm/GranberryAJ24} address a complementary
challenge -- \emph{what} to specify -- using symbolic methods to identify
under-specified clauses and focus LLM generation on the predicates that are actually missing. King et al.~\cite{DBLP:conf/icse-formalise/KingKK25} extend LLM-based generation to weakest preconditions and quantified array invariants, two classes that fall outside the scalar linear-arithmetic scope of most loop-invariant tools.
The paradigm also transfers to compiler settings: Magalhaes et al.~\cite{DBLP:conf/IEEEpact/MagalhaesWABJPO25} apply a \emph{guess, measure, and edit} feedback loop to guide LLM-based tensor code optimisation, demonstrating that structured neural--symbolic interaction is domain-agnostic.

\section{Conclusion}
\label{sec:conclusion}
Overall, we presented \ourtool{}, a neuro-symbolic pipeline that pairs locally-deployable open-weight language models with the \textsc{ESBMC} verification backend for automated loop invariant synthesis in C programs. The pipeline combines a deterministic symbolic phase with an iterative LLM refinement loop driven by structured verifier feedback, offering two prompting strategies: CoT and ToT. In our experimental evaluation, we show that cheap, locally hosted, open weight LLMs can be used effectively in a neuro-symbolic pipeline for loop invariant synthesis. \ourtool{} outperforms all other tools that run on local infrastructure and is competitive with LLM-based tools that rely on expensive external APIs and transmit source code to third-party services. Alongside this, our evaluation across five open-weight models, four inference strategies, and five benchmark families confirms that the \textsc{Basic} pipeline, which augments LLM refinement with symbolic invariant candidates, is the strongest strategy across all models. The symbolic phase resolves a significant fraction of problems without any LLM call and yields consistent gains for weaker models. Also, our ToT-inspired pipelines reveal a diversity--accuracy trade-off: ToT's multi-branch exploration reaches problem--model pairs inaccessible to direct CoT and yields a net gain for stronger models. In aggregate, however, non-ToT strategies dominate across most comparisons, as the iteration budget consumed by scouting outweighs the diversity benefit for weaker models, respectively. \ourtool{} matches the best prior LLM-augmented tools on shared benchmark suites and separates from them in generality -- handling the full SV-COMP loop set they cannot express -- while operating entirely on local infrastructure. A small fraction of problems remain unsolved across all pipelines and models, representing the hard frontier for future work.

\label{sec:future}

Several directions remain open for the future. First, for ToT, constrained decoding~\cite{willard2023efficient, beurer2023prompting} would reduce malformed outputs and free iteration budget for convergence, while ablating the scoring weights (Eq.~\ref{eq:score}) could shift the diversity--accuracy trade-off boundary across model tiers. On soundness, a post-processing pass that re-discharges accepted invariants under \textsc{ESBMC}'s bit-vector back-end would recover bit-precise semantics without burdening the LLM; extending Phase~1 candidate enumeration to array or separation-logic predicates~\cite{bradley2006s,reynolds2002separation} would open the pipeline to heap-manipulating programs. Finally, an energy and cost comparison against cloud-API baselines would empirically ground the efficiency claims that motivate local deployment, and a difficulty-aware early-stopping policy could make the iteration budget adaptive rather than fixed.

\begin{acks}
We thank Dr. Giles Reger for his advice, code assistance and guidance. We also thank Rafael Menezes for his support to create a setup to run the experiments. Julian Parsert was funded by the European Research Council (ERC) project LASD (Grant ID: 101089343).

The authors of this research disclose that generative AI was used in the development of the tooling and in the execution of the methodological evaluation. After using these tool(s)/service(s), the author(s) reviewed and edited the content as needed and take(s) full responsibility for the publication's content.
\end{acks}

\clearpage
\bibliographystyle{ACM-Reference-Format}
\bibliography{bibliography}

\appendix

\input{appendix/appendix}

\end{document}

%% file: content/system.tex
\ourtool{} is a neuro-symbolic pipeline for loop invariant synthesis (Figure \ref{fig:verIbmc}). Its design follows a single principle: \emph{escalate only when necessary}. Cheap, symbolic reasoning runs first, and the LLM is invoked only on the residue that symbolic methods cannot close. Notably, every candidate, regardless of its origin, is discharged by ESBMC before it is accepted. The pipeline is therefore sound by construction, and its LLM budget is spent exclusively on the problems that need it.

The pipeline comprises three phases. \emph{Phase~0} runs ESBMC on the unannotated program: if the program is trivially safe or provably unsafe, the pipeline exits immediately, so no synthesis effort is spent on problems the verifier already decides. \emph{Phase~1} symbolically enumerates candidate invariant atoms by pairwise comparison of loop-entry variables and constants, retaining those that ESBMC proves inductive (\prettyref{sec:symcand}). This phase is deterministic and requires no LLM call; it establishes a verified floor of invariant knowledge that every subsequent LLM interaction builds on. \emph{Phase~2} is the LLM refinement loop (\prettyref{sec:llm-int}): candidates are iteratively proposed, verified by ESBMC, decomposed into atoms, and recombined, with the verifier's verdicts fed back into subsequent prompts as structured evidence.

On top of this pipeline we define four inference strategies along two independent design choices (\prettyref{sec:strategies}). The \emph{verifier-feedback} axis controls whether the LLM receives Phase~1's symbolic prior: \textsc{Basic} runs all three phases, whereas \textsc{LLM-Only} skips Phase~1 and, within Phase~2, it also forgoes the atom-level decomposition, filtering, and recombination that \textsc{Basic} applies to every LLM proposal -- accepting a candidate only when it is inductive as a whole. \textsc{LLM-Only} thus serves as the ablation baseline that isolates raw LLM capability. The \emph{reasoning-structure} axis controls prompt organisation: each base strategy runs either with standard CoT prompting or with a ToT variant that explores four derivation styles in parallel (\prettyref{sec:tot-design}). \prettyref{sec:strategies} details the four strategies.




\subsection{ESBMC Loop Invariant Checker}
\label{sec:inv-checker}

Invariant synthesis and invariant checking are separate phases in VerIbmc. The checker is treated as a black-box primitive: for a program $P$ and a candidate invariant~$\varphi$ (a C Boolean expression placed at the loop head), it returns \textit{provable}, \textit{disprovable}, or \textit{unknown}. We realise this primitive using ESBMC's loop-invariant-checking mode\,\cite{esbmc_loop_invariants}.

\paragraph{Two operating modes.}
\ourtool{} invokes the checker in two structurally different ways. Let $P$ be the annotated program, $\psi$ its post-condition, and $\varphi$ a candidate invariant. 

\emph{(a) Atomic inductivity test.} To classify a single atom, the post-condition $\psi$ is stripped from $P$. $g$ is used as a guard which is
evaluated to true,and $\varphi$ is inserted as the only invariant. ESBMC's decision then reduces to the validity of
\begin{equation}
\label{eq:vc-atomic}
\mathit{VC}^{(a)}_\varphi \;\equiv\; \big(\mathit{pre}\rightarrow\varphi\big) \;\wedge\; \big(\varphi\wedge g\wedge\mathit{body}\rightarrow\varphi'\big),
\end{equation}
where $\mathit{pre}$ is the formula accumulated up to the loop head and $\varphi'$ renames each variable in $\varphi$ to its post-body version. A \textsc{successful} verdict witnesses that $\varphi$ is inductive; \textsc{failed} exhibits a concrete countermodel to one of the two conjuncts; timeout or parse error maps to \textit{unknown}.

\emph{(b) Whole-program verification.} To test whether a conjunction $\varphi=\bigwedge_i\varphi_i$, where each $\varphi_i$ is a provably-inductive atom, closes the proof, $\varphi$ is inserted and $\psi$ is kept; ESBMC then discharges both inductivity obligations together with the post-condition,
\begin{equation}
\label{eq:vc-whole}
\mathit{VC}^{(b)}_\varphi \;\equiv\; \mathit{VC}^{(a)}_\varphi \;\wedge\; \big(\varphi\wedge\neg g\rightarrow\psi\big).
\end{equation}
A \textsc{successful} verdict terminates the pipeline; \textsc{failed} indicates that $\varphi$ is inductive but too weak to entail $\psi$---a spurious counter-example in the sense of~\cite{10.1145/3691620.3695512}. \ourtool{} treats this as a signal rather than a defect: the atoms in $\varphi$ remain \textit{provable}, and recombination (\prettyref{sec:llm-int}) searches for additional atoms whose conjunction closes $\mathit{VC}^{(b)}_\varphi$. 



\subsection{\ourtool{} Phases}
As mentioned, \ourtool{} consists of 3 main phases. The first phases are simple symbolic reasoning while the last
is an iterative procedure that iteratively queries an LLM and subsequently filters and symbolically manipulates the feedback.
\subsection*{Phase~0: Baseline Verification}
\label{sec:phase0-method}
Before any invariant is generated, \ourtool{} uses ESBMC to verify the unannotated program~$P$. This ensures that we do not waste time on invariant synthesis when the verification conditions are trivial or ESBMC's standard methods are already strong enough to discharge them. In case of success or failure, we terminate immediately; on an unknown result, we proceed to the next phase. This procedure is deterministic and independent of model or strategy choice: its verdict depends only on ESBMC and the program.

\subsection*{Phase~1: Symbolic Candidate Generation}
\label{sec:symcand}
When Phase~0 returns \textsc{Unknown}, \ourtool{} synthesises invariant candidates by symbolically enumerating atoms without invoking an LLM. For each loop $\ell$ in $P$, let $V_\ell$ be the set of variables live at the loop head and $C_\ell$ the set of integer constants appearing in the loop body. We enumerate all \emph{atomic candidate atoms} of the form
\begin{equation}
v_i \bowtie v_j \quad v_i \bowtie c \quad \text{with} \quad \bowtie \in \{<,\leq,=,\neq,\geq,>\}
\end{equation}
Each atom is submitted to ESBMC in atomic-inductivity mode (Eq.~\ref{eq:vc-atomic}) and placed in one of three stores: $\mathcal{P}$ (\emph{provable}), $\mathcal{D}$ (\emph{disprovable}), or $\mathcal{U}$ (\emph{unknown}). The conjunction $\bigwedge\mathcal{P}$ is then submitted in whole-program mode (Eq.~\ref{eq:vc-whole}); if it closes the proof, the pipeline returns \textsc{Safe} without any LLM call. Otherwise, the populated stores are carried into Phase~2 as prior knowledge and grow monotonically as further atoms are synthesised and verified.

\paragraph{Worst-case complexity.}
For a loop with $|V|$ live variables and $|C|$ integer constants, Phase~1 enumerates at most
\begin{equation}
\label{eq:atoms}
N_{\text{atoms}} \;=\; 6\bigl(|V|^2 + |V|\cdot|C|\bigr)
\end{equation}
candidate atoms (the $|V|^2$ term covers all ordered variable--variable pairs, including $i{=}j$; self-comparisons such as $x \leq x$ are trivially true and filtered before ESBMC submission). Adding one extra relation symbol would scale the count by $7/6$; adding one constant adds $6|V|$ candidates. The enumeration is therefore linear in $|C|$ and quadratic in $|V|$, remaining tractable for the programs in our benchmark set; for multi-loop programs, Phase~1 runs independently per loop $\ell \in P$.

\begin{example}\label{ex:symbolic-generation}
The following example is benchmark from the code2inv set:
\begin{lstlisting}[language=C,basicstyle=\ttfamily\small]
int x = 1, y = nondet();
while (x <= 10) { y = 10 - x;  x++; }
assert(y < 10);
\end{lstlisting}
We have $V{=}\{x,y\}$ and $C{=}\{0,1,10\}$; Eq.~(\ref{eq:atoms}) gives at most $6(4+6)=60$ candidate atoms.
ESBMC classifies $x \geq 1$ and $x \geq 0$ as provable.
Atoms constraining $y$ alone---such as $y \geq 0$---are classified as \emph{disprovable}: since $y$ is nondeterministic at the first loop entry, no entry-point constraint on $y$ holds unconditionally.
The conjunction $\{x \geq 1,\, x \geq 0\}$ is checked but not strong enough to prove the assertion $y < 10$.
The pipeline enters Phase~2 with $\mathcal{P} = \{x \geq 1,\, x \geq 0\}$.
\end{example}
\subsection*{Phase~2: LLM Refinement Loop} 
\label{sec:llm-int}
In this phase \ourtool{} iteratively prompts an LLM for an invariant which we parse, check and in case of failure mutate to verify related invariants.
At each iteration, for every loop $\ell$ in~$P$, we construct a prompt that includes the loop's C source, the current contents of $\mathcal{P}$, $\mathcal{D}$, and $\mathcal{U}$ as feedback, and few-shot examples drawn from the prompting strategy (\prettyref{sec:prompts}). Note that the 3 stores that are populated are the only ``feedback'' we propagate from one iteration to the next.
We also experimented with including concrete ESBMC counterexamples directly in the prompt, but found this did not improve performance and often degraded it, particularly for smaller models. We attribute this to the well-documented tendency of LLMs to lose relevant information when the context grows long~\cite{liu2024lost, du2025context}: the additional counterexample trace increases prompt length without providing actionable signal beyond what the disprovable store $\mathcal{D}$ already encodes in a compact form. After submitting the prompt to the LLM we parse the response and for each invariant candidate $\varphi$ we:
\begin{enumerate}
  \item Attempt to verify the program using $\varphi$. In case of success, we terminate.
  \item In case of no success, we \emph{decompose} $\varphi$ into its sub-formulas and check individually (cf. \prettyref{eq:vc-atomic}) and add each to the appropriate store ($\mathcal{P}, \mathcal{D}, \mathcal{U}$).
  \item Finally, we use $\bigwedge\mathcal{P}$ as an invariant to verify the program where in case of success, we terminate with a successful verdict. Otherwise, we continue with the next iteration.
\end{enumerate}
In case of parsing errors, we simply continue with the next iteration. In order to improve the number of parsing errors
the prompt also contains instructions and examples of syntactically correct invariant candidates.
\begin{example}
Consider the program from \prettyref{ex:symbolic-generation}. 
Given $\mathcal{P}{=}\{x \geq 1,\, x \geq 0\}$ the LLM may generate the invariant:
\begin{lstlisting}[language=C,basicstyle=\ttfamily\small]
y == 0 && (x == 1 || y + x == 11)
\end{lstlisting}
Step~1: ESBMC attempts whole-program verification and \emph{fails}: the conjunct \texttt{y == 0} is not inductive since $y$ is nondeterministic at loop entry.
Step~2: \ourtool{} decomposes the formula into its conjuncts \texttt{y == 0} and \texttt{x == 1 || y + x == 11} and checks each individually. ESBMC classifies \texttt{y == 0} as \emph{disprovable} (added to $\mathcal{D}$) and \texttt{x == 1 || y + x == 11} as \emph{provable} (added to $\mathcal{P}$).
Step~3: The updated conjunction $\bigwedge\mathcal{P} = \{x{\geq}0,\, x{\geq}1,\, x{=}1{\vee}y{+}x{=}11\}$ is submitted for whole-program verification. Combined with the negated guard $x>10$, we obtain $x{\geq}11$ and $y{+}x{=}11$, hence $y = 11{-}x \leq 0 < 10$. This is sufficient to prove the assertion and \ourtool{} terminates successfully.
\end{example}

\subsection{Inference Strategies}
\label{sec:strategies}

\paragraph{Basic Strategy}
\label{sec:basic}
\textsc{Basic} chains all three phases with a single LLM. Phase~0 first checks whether ESBMC can decide the program unaided; if not, Phase~1 accumulates provable symbolic atoms and tests their conjunction. Only if that conjunction is insufficient does Phase~2 invoke the LLM, for up to $N$ iterations; each call contributes atoms to $\mathcal{P}$, which feed forward as context into subsequent prompts. The LLM is thus never asked to rediscover what the symbolic phase has already established: it starts from a verified partial invariant and is steered toward the atoms that are still missing.

\paragraph{LLM-Only Strategy}
\label{sec:llm-only}
\textsc{LLM-Only} omits Phase~1: after Phase~0, the pipeline enters Phase~2 with empty stores. The provable/disprovable/unknown lists are used only for deduplication, preventing re-submission of previously evaluated candidates, and carry no symbolic prior into the prompt. This strategy is structurally analogous to \textsc{ESBMC-ibmc}~\cite{10.1145/3691620.3695512} and serves as the primary ablation baseline: any gap between \textsc{Basic} and \textsc{LLM-Only} is attributable to Phase~1's symbolic augmentation.

\paragraph{ToT Variants}
\label{sec:tot-variants}
Each base strategy has a ToT counterpart (\textsc{Basic-ToT}, \textsc{LLM-Only-ToT}) that replaces single-strategy prompting in Phase~2 with a two-stage multi-branch search: a scouting stage issues one LLM call per derivation style (\prettyref{sec:prompts}), scores each style by the inductive atoms it yields, and a refinement stage concentrates the remaining iteration budget on the top-two styles. The verifier-feedback axis is unaffected: \textsc{Basic-ToT} retains the Phase~1 prior, \textsc{LLM-Only-ToT} does not. The search algorithm and scoring function are detailed in \prettyref{sec:tot-design}.

\subsection{Prompt Engineering}
\label{sec:prompts}
In the default case, we use a few-shot CoT prompt containing a task instruction, the target C program, the current invariant store as feedback, and one of four \emph{example sets} demonstrating different derivation styles. These prompt strategies are:
\begin{description}
  \item[Inductive with Explanation.]
  Each example shows the program, the invariant annotation, and a step-by-step inductive proof (base case, inductive step, and sufficiency check).
  This style targets models that benefit from explicit reasoning traces.
  \item[Inductive without Explanation.]
  The same programs and annotations are shown but without derivation text.
  This compact style reduces prompt length and is suited for models sensitive to context size.
  \item[Hoare Logic Derivation.]
  Each example derives the invariant by constructing and discharging the Hoare triple $\{I \wedge B\}\;S\;\{I\}$, making the consecution obligation explicit.
  This style anchors the model's generation in the formal semantics of the loop.
  \item[Horn Clause Derivation.]
  Each example encodes the loop as a constrained Horn clause system and derives the invariant by logical consequence.
  This style is particularly effective for non-linear arithmetic invariants.
\end{description}
In the basic strategy, a single example set is selected for the entire run.
In the ToT variants, all four sets are explored in parallel and the best two are selected for refinement (\prettyref{sec:tot-design}).
\subsection{ToT Refinement}
\label{sec:tot-design}

Firstly, let $\pi_s(\mathcal{I})$ denote the prompt template for strategy $s$, augmented with the current set of verified invariant atoms $\mathcal{I}$. Next, let $\pi^{\mathrm{fb}}_s(\mathcal{I})$ denote the feedback-augmented prompt for strategy $s$, which extends $\pi_s(\mathcal{I})$ with the counterexample and verification outcome from the previous LLM candidate.

\begin{algorithm}
\caption{Invariant Synthesis using ToT prompting.}
\label{alg:tot}
\begin{algorithmic}[1]

\Require $P$, $M$, strategies $\mathcal{S}$, $R{=}10$, $T$
\Ensure Verification result

\State \textbf{if} $\textsc{Vfy}(P,\emptyset){=}\textsc{Succ}$ \textbf{then} \Return \textsc{Success}
\State \textbf{if} $\textsc{Vfy}(P,\emptyset){=}\textsc{Fail}$ \textbf{then} \Return \textsc{Unsolvable}
\State $\mathcal{I}\gets\emptyset$

\ForAll{$s\in\mathcal{S}$} \Comment{Phase 1: one attempt per strategy}
    \State $C\gets\textsc{LLM}(M,\,\pi_s(\mathcal{I}))$
    \State \textbf{if} $C$ verifies \textbf{then} \Return \textsc{Success}
    \State add provable atoms of $C$ to $\mathcal{I}$
    \State \textbf{if} $\mathcal{I}$ verifies \textbf{then} \Return \textsc{Success}
    \State $\mathrm{score}(s)\gets(0.6|\mathcal{P}_s|-0.3|\mathcal{D}_s|+0.1|\mathcal{U}_s|)/|\mathcal{P}_s\cup\mathcal{D}_s\cup\mathcal{U}_s|$
    \State \textbf{if} $\textsc{Elapsed}()\geq T$ \textbf{then} \Return \textsc{Timeout}
\EndFor

\State $\mathcal{S}^*\gets\mathrm{top}_2(\mathcal{S})$;\;$r\gets0$;\;$\mathrm{step}[s]\gets0$

\While{$r<R$ \textbf{and} $\mathcal{S}^*\neq\emptyset$} \Comment{Phase 2: top-2 refinement}
    \State $s\gets\mathcal{S}^*[r\bmod|\mathcal{S}^*|]$
    \If{$\mathrm{step}[s]\geq K_s$}
        \State $\mathcal{S}^*\gets\mathcal{S}^*\setminus\{s\}$;\,\textbf{continue}
    \EndIf
    \State $r\mathrel{+}=1$;\;$\mathrm{step}[s]\mathrel{+}=1$
    \State $C\gets\textsc{LLM}(M,\,\pi^\mathrm{fb}_s(\mathcal{I}))$
    \State \textbf{if} $C$ verifies \textbf{then} \Return \textsc{Success}
    \State add provable atoms of $C$ to $\mathcal{I}$
    \State \textbf{if} $\mathcal{I}$ verifies \textbf{then} \Return \textsc{Success}
    \If{$r\bmod6{=}0$}
        \State $\mathcal{S}^*\gets\mathrm{top}_2(\mathcal{S})$;\;$\mathrm{step}[\cdot]\gets0$ \Comment{Re-rank every 6 rounds}
    \EndIf
    \State \textbf{if} $\textsc{Elapsed}()\geq T$ \textbf{then} \Return \textsc{Timeout}
\EndWhile

\State \Return \textsc{Failure}

\end{algorithmic}
\end{algorithm}
We hypothesise that ToT is well-suited to loop invariant inference for multiple reasons. First, the space of candidate invariants is large and non-monotone: a partial invariant that is not provable become valid once it is weakened. Second, the symbolic verifier provides a \emph{structured} evaluation signal, which can be used directly as the state evaluator to prune branches early. Third, multiple candidate invariants can be checked in parallel using independent SMT solver processes, aligning naturally with ToT's multi-branch search structure. These properties motivate the design of our ToT-based prompting strategy.

\prettyref{alg:tot} receives: program~$P$; model~$M$; strategy set $\mathcal{S}=\{S_1,S_2,S_3,S_4\}$ corresponding to the four CoT example sets (\prettyref{sec:prompts}); maximum refinement rounds $R{=}10$; and timeout~$T$.
The goal is to identify---via a single scored pass over all four strategies in Stage~1---which prompting style generates the most inductive atoms for the given program, then concentrate the remaining iteration budget on the top-2 strategies in Stage~2 rather than committing blindly to one from the start.
It proceeds in two stages.

\paragraph{Stage~1 (strategy scoring).}
One LLM call is made per strategy $s \in \mathcal{S} = \{S_1, S_2, S_3, S_4\}$, each using its corresponding example set.
For each strategy~$s$, let $\mathcal{P}_s$, $\mathcal{D}_s$, $\mathcal{U}_s$ be the provable, disprovable, and unknown atom sets accumulated from $s$'s Stage~1 response. A score is computed:
\begin{equation}
\label{eq:score}
  \mathrm{score}(s) \;=\; \frac{0.6\,|\mathcal{P}_s| \;-\; 0.3\,|\mathcal{D}_s| \;+\; 0.1\,|\mathcal{U}_s|}{|\mathcal{P}_s \cup \mathcal{D}_s \cup \mathcal{U}_s|}.
\end{equation}
Provable atoms carry the most signal toward a successful proof; disprovable atoms are penalised because they contribute no usable invariant content; unknown atoms receive a small positive weight to encourage exploration of strategies that produce candidates the verifier cannot yet classify.
A strategy achieving whole-program verification short-circuits with $\mathrm{score}(s)=1.0$; a strategy producing no atoms receives a fallback score of $0.1$.


Traditionally, ToT prompting asks the LLM to self-evaluate and rank the strategies~\cite{yao2023tree}.
However, in our implementation the actual top-2 selection in \prettyref{alg:tot} uses the ESBMC-derived score (Eq.~\ref{eq:score}), not the LLM's self-assessment: the LLM cannot run the verifier and therefore lacks ground truth to self-evaluate reliably.
Although we override the LLM's preference with the ESBMC-derived score (Eq.~\ref{eq:score}), we retain the self-ranking instruction in the prompt.
Asking the model to compare strategies and commit to a preference \emph{before} generating candidates acts as a think-before-generate step: empirically, this structured prompt discipline produces more coherent and more frequently inductive invariant proposals than unstructured generation, even though the expressed preference itself is not used by the search controller.

\paragraph{Stage~2 (top-2 refinement).}
The two highest-scoring strategies proceed to a refinement loop of up to $R{=}10$ rounds, cycling between the two.
Every 6~rounds the scores are recomputed from the full accumulated evidence and the top-2 selection is refreshed, allowing a previously lower-ranked strategy to re-enter.
This interval was chosen so that, within the $R{=}10$ round budget, each selected strategy accumulates at least three rounds of evidence before re-evaluation.
The per-strategy counter $\mathrm{step}[s]$ caps each strategy's contribution at $K_s = \lfloor R/|\mathcal{S}^*| \rfloor$ rounds (i.e.\ an equal share of the total budget); exhausted strategies are removed from $\mathcal{S}^*$.

%% file: content/experiments.tex
\section{Experimental Evaluation}
\label{sec:results}
\label{sec:evaluation}
We conduct an experimental evaluation of \ourtool{} to answer the following research questions:
\begin{description}
    \item[RQ1:] Can symbolic methods combined with locally deployed open-weight models effectively synthesise loop invariants?
    \item[RQ2:] Does this neuro-symbolic method advance the state of the art in loop invariant synthesis?
    \item[RQ3:] What effect does Tree-of-Thoughts prompting have on the efficacy of neuro-symbolic invariant synthesis?
    \item[RQ4:] How do the individual pipeline phases contribute to the overall solve rate?
\end{description}

The experiments involve five LLMs, four inference strategies, and eleven benchmark suites, totalling 10\,400 per-problem outcomes. A supplementary dual-LLM experiment is reported separately in Appendix~\ref{sec:dual-llm-supp}. The code is available at: https://zenodo.org/records/20690105.
We also conduct experiments to compare with other tools that represent the state-of-the-art (SOTA). These consist of the following: ESBMC \cite{gadelha2019esbmc, Sa_Menezes_ESBMC_7_4_Harnessing}, CPAchecker \cite{beyer2011cpachecker}, 2ls \cite{schrammel20162ls}, LaM4Inv \cite{DBLP:conf/kbse/WuC0W0M24}, Clause2Inv \cite{DBLP:journals/pacmse/CaoWXYWCM25}, Code2Inv \cite{si2018learning}, and LORIS~\cite{10.1145/3806652}.

\subsection{Experimental Setup}
\label{sec:setup}

\paragraph{Benchmark suites.}
We evaluate on five benchmark families totalling 520 problems. Four are established loop-invariant synthesis suites: \emph{C2I} (133 problems), introduced with Code2Inv~\cite{si2018learning}; \emph{L4I-SVC} (99) and \emph{L4I-SY} (84), introduced with LaM4Inv~\cite{DBLP:conf/kbse/WuC0W0M24} and drawn from SV-COMP~2024 and SyGuS~2019 respectively; and \emph{NL} (50), the non-linear suite introduced with Clause2Inv~\cite{DBLP:journals/pacmse/CaoWXYWCM25}. The fifth family, \emph{SVC} (154), is our own selection from the latest SV-COMP edition~\cite{DBLP:conf/tacas/BeyerS26}: the complete \texttt{loops}, \texttt{loops-crafted-1}, \texttt{loop-acceleration}, \texttt{loop-crafted}, \texttt{loop-invariants}, \texttt{loop-new}, and \texttt{loop-\newline{simple}} subcategories. Unlike the four established suites, which were syntactically restricted to fit the front-ends of the tools that introduced them, the SVC family is unfiltered: it retains programs with arrays, nested and multiple loops, pointers, and function calls.

Of the 520 problems, 21 are unsolvable under machine-integer semantics (ESBMC reports \texttt{Verification Failed} due to unavoidable overflow), leaving an \emph{effective} pool of 499 problems. All solve rates in this section use 499 as the denominator; RQ4 further decomposes this pool by pipeline phase.

\paragraph{Hardware and software.}
All models are served locally via Ollama on a server running Ubuntu 22.04.5 LTS and equipped with four NVIDIA RTX A6000 GPUs (48\,GB each). ESBMC v8.2 serves as the backend verifier throughout all experiments.

\paragraph{Models.}
We focus on open-weight models that can be deployed entirely on local infrastructure, avoiding the need to transmit source code to external services. We evaluate five models spanning 7B to 120B parameters: GPT-OSS-120B \cite{agarwal2025gpt}, GPT-OSS-20B \cite{agarwal2025gpt}, Qwen2.5-32B-Instruct \cite{qwen2025qwen25}, Qwen2.5-7B-Instruct \cite{qwen2025qwen25}, and Llama-3.1-8B \cite{grattafiori2024llama}. The GPT-OSS models use their default temperature of 1.0, while the remaining models use Ollama's default temperature setting of 0.8.

\paragraph{Inference strategies.}
We evaluate four inference strategies built from two independent design choices. The first is verifier feedback: the \emph{LLM-Only} strategy enters Phase~2 directly after Phase~0 with empty atom stores, so the LLM receives no symbolic prior and each LLM proposal is accepted only if it is inductive as a whole; the \emph{Basic} strategy runs the full three-phase pipeline (Figure~\ref{fig:verIbmc}), augmenting each LLM call with the provable atoms accumulated in Phase~1 and, in addition, decomposing every LLM proposal into atoms, discarding the non-inductive ones, and recombining the provable remainder into the strongest verified invariant. LLM-Only therefore differs from Basic along two coupled mechanisms---the symbolic prior and this atom-level filtering/recombination of LLM output---rather than the symbolic prior alone, and serves as the primary ablation baseline isolating raw LLM capability from symbolic augmentation. The second design choice is reasoning structure. Each of the two strategies above can be run with standard CoT prompting or with a ToT variant, denoted by the suffix \texttt{-tot}, which explores all four prompting styles in parallel before concentrating the iteration budget on the top-two (Section~\ref{sec:tot-design}). This yields four strategies in total: \emph{LLM-Only}, \emph{Basic}, \emph{LLM-Only-ToT}, and \emph{Basic-ToT}.

\paragraph{Experimental protocol.}
Each problem is run with a 600\,s wall-clock timeout.
Outcomes are classified as success, timeout, failure, or unknown. We further distinguish normal successes from \emph{integer-relaxed} (IR) successes: an IR success means ESBMC verifies the property only when integer widths are modelled as unbounded rather than fixed-width machine integers. The \textbf{Total} column and all headline solve rates include IR solves. The IR variation across configurations is small (at most 14 out of 499 problems) and does not alter any qualitative ranking in Sections~\ref{sec:rq1}--\ref{sec:rq3}. The SOTA comparison (Section~\ref{sec:rq2}) includes IR solves to match the unbounded-integer semantics assumed by all competing tools (see Section~\ref{sec:threats}). When we initially ran \ourtool{}, our best performing strategy was \textsc{Basic} with the LLM \textsc{GPT-OSS-120B}. Since LLM responses are subject to non-determinism, we ran this strategy three times and report the best-performing run (see Table~\ref{tab:reproducibility}), following the established \emph{pass@k} evaluation protocol used in LLM-based code-generation benchmarks~\cite{chen2021evaluating}. The variance across the three runs was small (see Table~\ref{tab:reproducibility}), confirming that best and average would yield equivalent qualitative rankings.


\subsection{Results}

\subsubsection*{RQ1: Can symbolic methods combined with locally deployed open-weight models effectively synthesise loop invariants?}
\label{sec:rq1}

\begin{table}
\centering
\input{results_table_cot}
\caption{VerIbmc results across 520 benchmark problems (499 effective). Column headers: \textbf{C2I} = Code2Inv, \textbf{L4I-SVC} = LaM4Inv-SVCOMP24, \textbf{L4I-SY} = LaM4Inv-SyGuS19, \textbf{SVC} = SV-COMP (full), \textbf{NL} = Non-Linear. \textbf{Total} = sum of all five benchmark columns plus the \textbf{IR} column (problems solvable only under unbounded-integer semantics). The phase-level ablation in Table~\ref{tab:ablation} (Appendix) reports only standard-semantics solves; the difference equals the IR column. Basic results are the best of three independent runs per model (see Table~\ref{tab:reproducibility})}
\label{tab:results_veribmc_cot}
\end{table}

We examine overall solve rates across all (model, strategy) pairs, isolating the contribution of each pipeline phase, and characterising convergence behaviour. The results of the standard (non-ToT) strategies are shown in \prettyref{tab:results_veribmc_cot}. The Basic strategy is the best-performing for every model, achieving 1{,}931 solved instances across 2{,}600 model--problem pairs (74.3\%).
For the strongest model, GPT-OSS-120B, Basic solves 431/499 (86.4\%) and outperforms LLM-Only (428/499). The GPT-OSS-20B model solves 424 problems, only 7 problems fewer than the stronger model. In contrast, the other models are significantly weaker with Qwen2.5-32B solving 382 problems. Across all models, the Basic pipeline matches or outperforms the LLM-Only pipeline, indicating that the addition of symbolic reasoning is advantageous.
This effect is most pronounced when using Llama-3.1-8b and Qwen2.5-7B, indicating that symbolic reasoning can mitigate weaker-model performance.
Across benchmark families, performance varies considerably: \emph{C2I} is solved to near-completion under Basic for all five models, whereas \emph{L4I-SY} and \emph{NL} are more discriminating. The strongest model (GPT-OSS-120B) shows its largest relative advantage on the non-linear \texttt{clause2inv\_NL} suite, where generating polynomial invariants is beyond the reach of weaker models.
In conclusion, the experiments indicate that \ourtool{} can be used for effective invariant synthesis. Furthermore, the choice of LLM has a significant impact on the results, and the use of symbolic invariants has a positive impact, especially on weaker models.
%
%
%
\subsubsection*{RQ2: Does this neuro-symbolic method advance the state of the art in loop invariant synthesis?}
\label{sec:rq2}\label{sec:sota}
To answer this question, we compare \textsc{VerIbmc} against three classes of tools, summarised in \prettyref{tab:results_sota}.
\emph{Class~I (traditional symbolic):} CPA-Checker, 2ls, and ESBMC-Kind.
\emph{Class~II (neural):} Code2Inv~\cite{si2018learning}, which learns loop invariants via graph neural networks.
\emph{Class~III (LLM-augmented):} Clause2Inv, LaM4Inv, and LORIS.
All Class~I and II tools were run directly on our hardware across all benchmarks, enabling a direct and fair comparison.
\textsc{VerIbmc} substantially outperforms every tool in these two classes: the best single configuration (GPT-OSS-120B, 431/499; 86.4\%) and the overall union across the four single-LLM strategies (All LLMs, 437/499, 87.6\%) is significantly more than the state-of-the-art verifiers, e.g., CPA-Checker's 282 solved problems, with Code2Inv at 210, 2ls at 194, and ESBMC-Kind at 145. On the SV-COMP benchmarks \ourtool{} solves 85 and 89 problems in the IR configuration while CPA-Checker solves 76 problems, 2ls 57 and ESBMC-Kind 43.
\begin{table*}
\centering
\input{sota_tool_table}

\caption{Comparison with state-of-the-art tools. \textbf{C2I} = Code2Inv,
\textbf{L4I-SVC} = LaM4Inv-SVCOMP24, \textbf{L4I-SY} = LaM4Inv-SyGuS19,
\textbf{NL} = Non-Linear. \textbf{Subtotal$^\ddagger$} sums the four shared suites. \textbf{Total$^\dagger$} = Subtotal $+$ Full\,(154); the Comp.\,(43) column is a subset of Full\,(154) and is therefore excluded from the sum. Tools whose frontend cannot run the full SV-COMP set have no Full\,(154) figure and hence no Total ($^*$). $^*$~frontend-limited; see Table~\ref{tab:svc-limitation}.
Comp.\,(43) entries for ESBMC-Kind, 2ls, CPA-Checker, and Code2Inv are from our own runs on the comparable subset (same 600\,s timeout). LORIS uses GPT-4.1; \ourtool{} uses local models. VerIbmc = best single (model, strategy) cell; VerIbmc* = union across all configurations.}
\label{tab:results_sota}
\end{table*}

Class I and II tools, as well \ourtool{} require no external API subscriptions and no transmission of source code to external servers. In contrast, the Class~III LLM-augmented tools rely on external paid APIs (e.g. OpenAI or Anthropic) that cannot be fully replicated in our setup, requiring a more nuanced comparison. We therefore report their published figures for the four shared benchmark suites (\emph{C2I}, \emph{L4I-SVC}, \emph{L4I-SY}, \emph{NL}).
Furthermore, their provided implementations are subject to input-format constraints: Clause2Inv and LaM4Inv use a \texttt{clang-fe} frontend that encodes only single-loop scalar integer programs, and LORIS's front-end enforces the same restriction; comparison on the SV-COMP problem set is therefore restricted to the 43-program subset that their respective parsers can handle.
Additionally, on the shared benchmark suites, \ourtool{} (346--348 solved, union across the four single-LLM strategies) outperforms 
LaM4Inv (338), and is competitive with -- though slightly below -- Clause2Inv (356) and LORIS (351). 
Note that these Clause2Inv and LORIS both require state of the art expensive frontier models run on external servers.
On the comparable SV-COMP subset, LORIS (GPT-4.1) solves $39/43$, \ourtool{} solves $36/43$ using local models, and Clause2Inv and LaM4Inv each solve $30/43$; pure ESBMC solves only $5/43$ without invariants, CPA-Checker verifies $19/43$, 2ls verifies $14/43$ and Code2inv verifies $11/43$, confirming the gain is due to invariant synthesis.
\ourtool{} with GPT-OSS-20B (and even an 8B Llama at $33/43$) stays at or above Clause2Inv's and LaM4Inv's $30$, despite their use of GPT-4o-mini.

A breakdown of the different problem sets shows that the best single configuration solves all 133 \emph{C2I} problems (matching LaM4Inv's perfect score), 94/99 on \emph{L4I-SVC} (within one of LaM4Inv's 95, four below Clause2Inv's 98), 78/84 on \emph{L4I-SY} (three below LaM4Inv's 81, four below Clause2Inv's 82), and 41/50 on \emph{NL} (three below Clause2Inv's 44, four below LORIS's 45). Note that the Class I and II tools are clearly beaten by a large margin in every problem set.
Non-linear arithmetic is a known weak point of SMT solvers, the best  Class I tool solves only 8 problems here, so LLM-generated invariants serve as guided lemma selection, enabling ESBMC to discharge polynomial goals it would otherwise fail on.

In summary, \ourtool{} substantially outperforms all Class~I and II tools, is competitive with the best LLM-augmented tools on the benchmarks they can express, and uniquely scales to the full range of C programs. Notably, in our experiments Class I and II tools were run on the same hardware as \ourtool{} and without any calls to external servers.
%
%
%

\subsubsection*{RQ3: What effect does Tree-of-Thoughts prompting have on the efficacy of neuro-symbolic invariant synthesis?}
\label{sec:rq3}
\begin{table}
\centering
\input{results_table_tot}
\caption{Results for the ToT strategies. Column abbreviations and IR convention as in \prettyref{tab:results_veribmc_cot}. Shaded cells indicate where ToT outperforms its non-ToT counterpart in aggregate; ToT additionally reaches exclusive problems not reflected in these totals (\prettyref{sec:rq3}).}
\label{tab:results_veribmc_tot}
\end{table}
The results of ToT based pipelines are shown in \prettyref{tab:results_veribmc_tot}.
Comparing \prettyref{tab:results_veribmc_cot} with \prettyref{tab:results_veribmc_tot} reveals that the performance of ToT hinges on the strength of the model.
For strong models the difference to the CoT prompting is negligible while for the weaker models
the performance is lacking.
\begin{table}
\centering
\input{tot_comparison_table}
\caption{CoT versus ToT solve counts and total wall-clock time per model and strategy.
Numbers in parentheses are total wall-clock time across all 11 benchmark families.
$\Delta$ = ToT $-$ CoT (solve count).
\emph{ToT-unique}: problems where the model fails under both Basic and LLM-Only
but succeeds under at least one ToT strategy.}
\label{tab:tot-comparison}
\end{table}
Table~\ref{tab:tot-comparison} summarises the CoT--ToT comparison.
ToT's key contribution is not aggregate throughput but \emph{diversity}: its scouting stage explores multiple candidate invariants and reaches 49 (model, problem) pairs that the direct CoT approach cannot, two of which are globally unique across all models and strategies.
The aggregate cost of this diversity depends on model capability: for the two strongest models the deltas are small ($-5$ to $-13$), with GPT-OSS-20B under LLM-Only-ToT the sole case of a net gain ($+5$).
Weaker models pay a larger price, most severely Llama-3.1-8B ($-57$ under Basic-ToT).
In total, 119 pairs move in the opposite direction under ToT; the failure-mode decomposition in Section~\ref{sec:discussion} explains why.

Table~\ref{tab:tot-comparison} also shows that ToT incurs a wall-clock overhead over its base counterpart, driven by the scouting stage; this ranges from negligible (Llama-3.1-8B Basic-ToT, $+2\%$) to $1.40\times$ for LLM-Only-ToT on Qwen2.5-7B (5.7\,h $\to$ 8.0\,h). Within the non-ToT strategies, Basic and LLM-Only run in comparable time, confirming that the extra work Basic performs---Phase~1 symbolic enumeration together with the atom-level filtering and recombination of each LLM proposal---is negligible relative to the LLM loop, so its accuracy gains in RQ1 come effectively for free.

\subsubsection*{RQ4: How do the different pipeline phases contribute to the overall solve rate?}
\label{sec:feedback}\label{sec:phase0}
\begin{figure*}
\centering
\begin{subfigure}[t]{0.5\textwidth}\centering
\includegraphics[width=\linewidth]{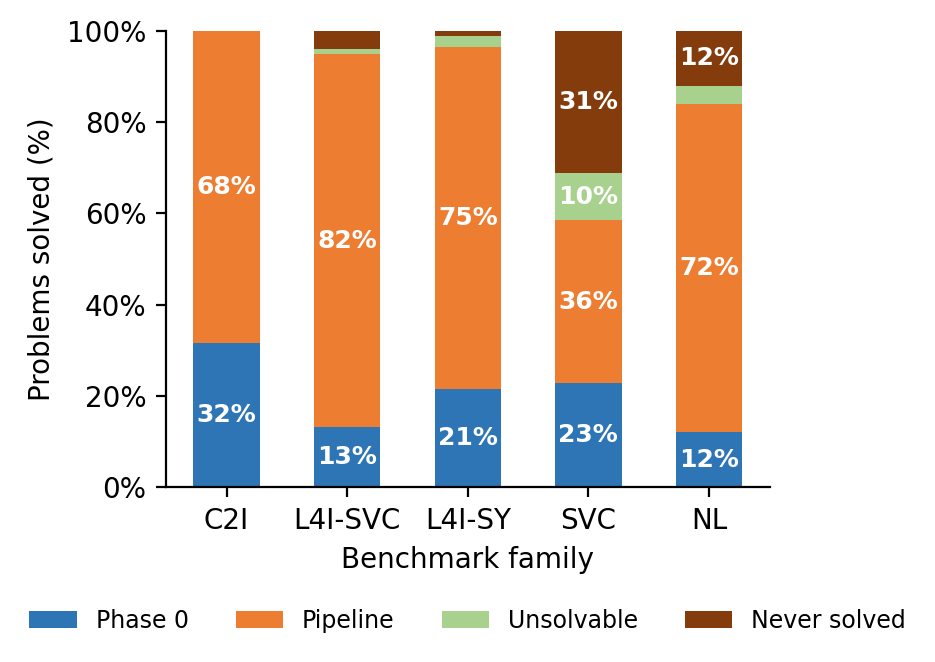}
\caption{Per-benchmark breakdown of solved contribution by phase (as a percentage of each family's total \newline{problems}).}
\label{fig:phase0-bar}
\end{subfigure}
~
\begin{subfigure}[t]{0.5\textwidth}
    \centering
\includegraphics[width=\linewidth]{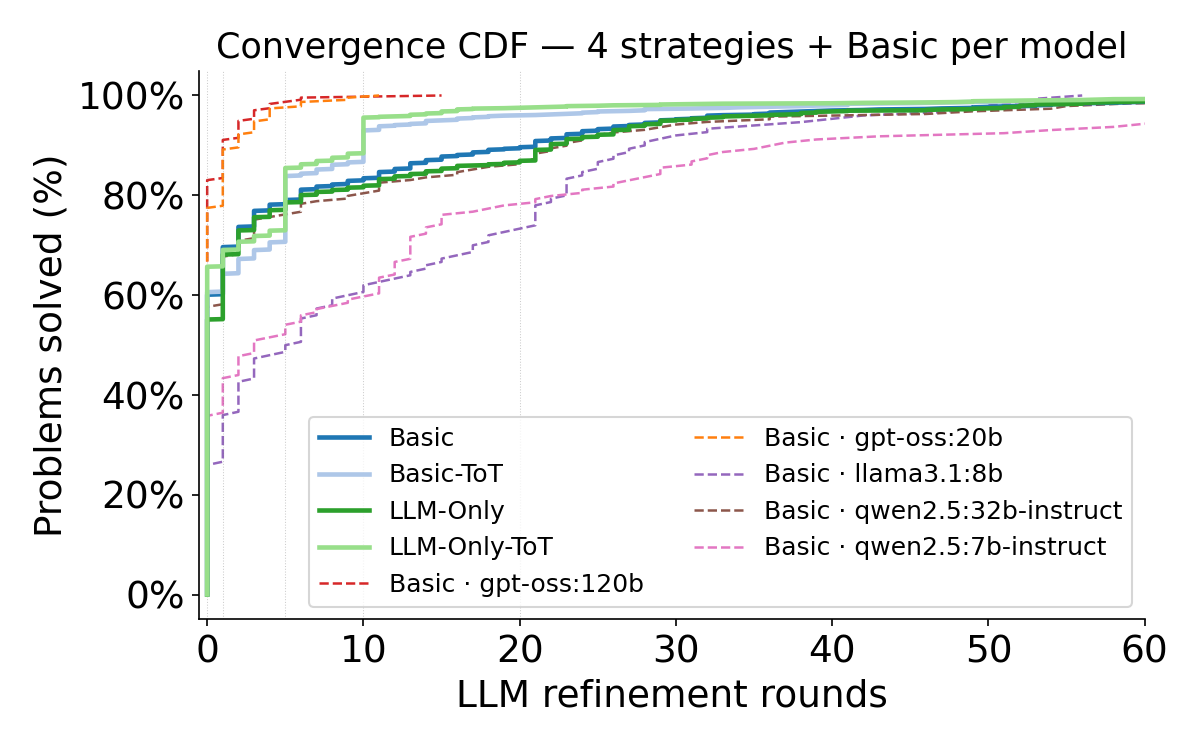}
\caption{Cumulative Distribution Function (CDF) of refinement rounds on successful Phase~2 outcomes, broken down by model. Each curve aggregates all four main strategies for that model. }
\label{fig:itercdf}
\end{subfigure}
\caption{Solve rates by refinement iterations and benchmark family.}
\end{figure*}

To see which phases had what impact on the solve rate we look at the solved problems and look at the which phase of the pipeline in \prettyref{fig:verIbmc} led to a successful verification.
Overall, the counts are as follows:
\begin{itemize}
  \item \textbf{111 (22.2\%)} are solved by Phase~0 alone under standard semantics, plus \textbf{3 (0.6\%)} under unbounded-integer (IR) semantics -- \textbf{114} Phase~0 solves in total;
  \item \textbf{75 (15.0\%)} are solved by symbolic stage in Phase~1;
  \item \textbf{248 (49.7\%)} need at least one LLM-generated invariant phase; 
  \item \textbf{62 (12.4\%)} remain unsolved by every (model, strategy) combination within the 600\,s budget.
\end{itemize}
Figure~\ref{fig:phase0-bar} visualises the breakdown per benchmark family.
Phase~0 is strongest in \emph{C2I}; \emph{SVC} shows the highest proportion of never-solved problems, making it the hardest family.
Phase~1 symbolic candidates contribute a small but fully deterministic share across most suites. The 323 problems requiring the LLM/symbolic stage are therefore the right denominator for evaluating the LLM's contribution---though even within this subset, 75 are resolved by Phase~1's symbolic candidates alone, without any LLM call.

On this 323-problem subset, performance diverges sharply by model: GPT-OSS-120B/Basic solves 314 (97.2\%), Llama-3.1-8B/Basic solves 229 (70.9\%), and Llama-3.1-8B/Basic-ToT solves only 172 (53.3\%).

Phase~1 symbolic candidates alone resolve exactly \textbf{75} problems that Phase~0 cannot decide---a fully deterministic contribution independent of which LLM is paired with the system.
Table~\ref{tab:ablation} (Appendix) shows the per-phase breakdown for each model.
Llama-3.1-8B gains a net \textbf{+36} problems (75 Phase~1 solves, offset by $-39$ LLM solves on the harder residual); Qwen2.5-7B gains \textbf{+22}; GPT-OSS-20B gains \textbf{+15}; GPT-OSS-120B gains zero, as the strong model independently derives the same provable atoms.

%
%
%
\subsection{Discussion}
\label{sec:discussion}

\paragraph{Convergence behaviour.}
Figure~\ref{fig:itercdf} shows the distribution of refinement rounds required to solve each problem.
Convergence is heavily front-loaded across all models and strategies: 65.1\% of LLM-solved problems are solved on the first iteration, 85.1\% within five, and 88\% within ten; the maximum is 171 iterations.
Unsurprisingly, stronger models (GPT-OSS-120B, GPT-OSS-20B) converge faster, with their curves reaching the plateau at lower iteration counts, while weaker models (Llama-3.1-8B, Qwen2.5-7B) show a longer tail.
This front-loading suggests that most problems are either directly solvable by the model or require only modest refinement; the long tail (rounds $>20$) accounts for fewer than 10\% of successful solves and represents structurally harder instances where the LLM must progressively accumulate provable atoms before closure.

For GPT-OSS-120B, Basic (431/499) marginally outperforms LLM-Only (428/499) by 3 problems. Phase~1 symbolic candidates primarily \emph{displace} LLM solves rather than add new ones for the strong model: it independently derives the same provable atoms through direct generation that Phase~1 enumerates symbolically, explaining why the scores are nearly identical. Only a small number of problems genuinely require Phase~1 to succeed where LLM generation alone does not converge within budget.

\begin{figure}
\centering
\includegraphics[width=\linewidth]{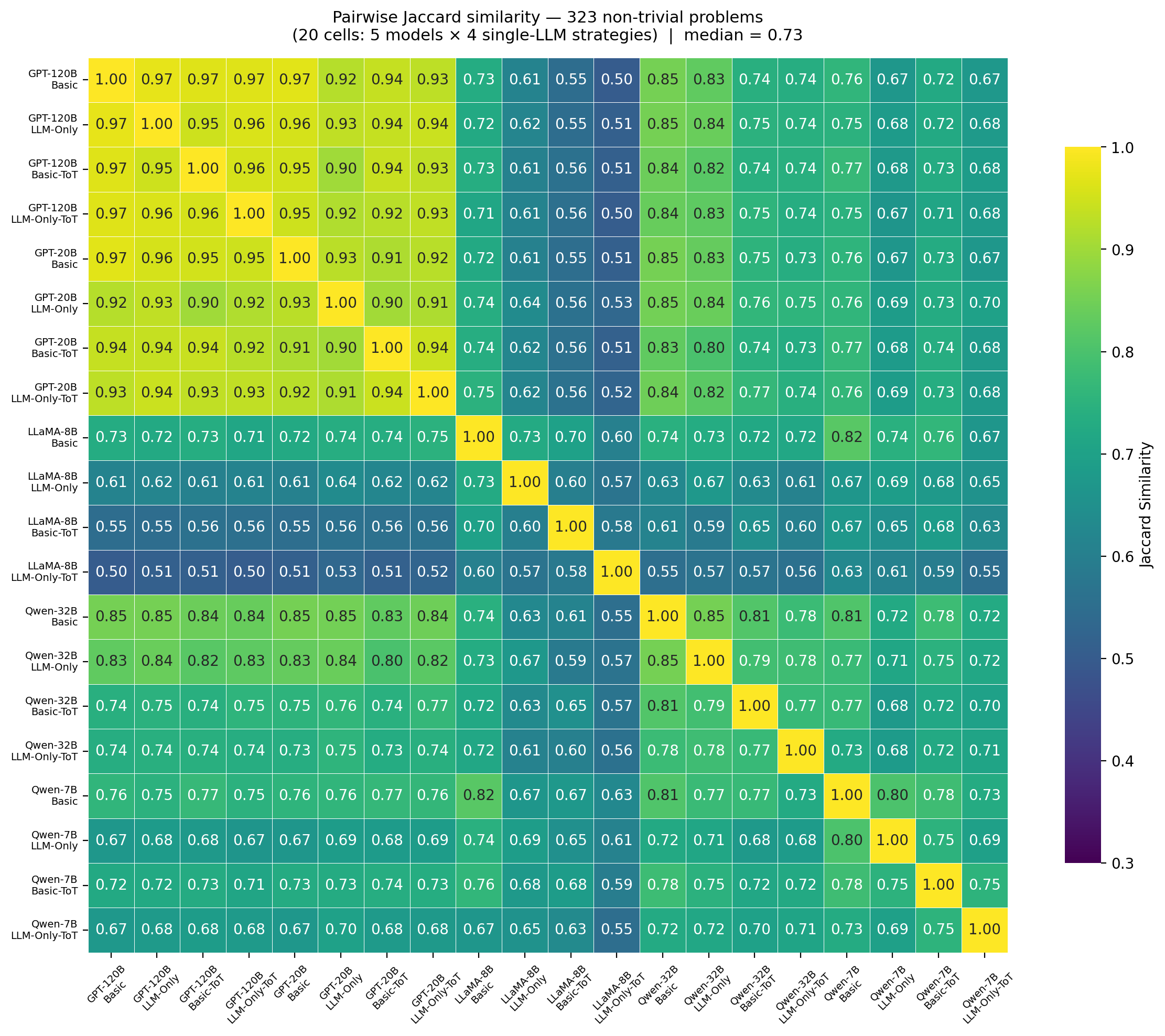}
\caption{Pairwise Jaccard similarity of solved-problem sets across the 20 (model, strategy) cells of our evaluation (5 models $\times$ 4 strategies), restricted to the 323 non-trivial problems. Median pairwise Jaccard $= 0.73$. Strong off-diagonal agreement ($\geq 0.7$ in nearly every cell) indicates that the solvable frontier is robust to model/strategy choice once the Phase~0 baseline is held constant.}
\label{fig:disagreement}
\end{figure}

\paragraph{Unique Solves.}
The four strategies clusters cover between 314 and 318 of the 323 non-trivial problems, and at most 1 problem is unique to any single strategy.
Figure~\ref{fig:disagreement} renders the per-cell pairwise Jaccard similarity on the non-trivial subset across the 20 (model, strategy) cells; median pairwise Jaccard is $\approx 0.73$, confirming strong agreement.
Each entry $(i,j)$ reports $|S_i \cap S_j|/|S_i \cup S_j|$; a value of 1.0 denotes identical coverage and 0.0 denotes no overlap.
Pairs sharing the same model (different strategy) exhibit the highest Jaccard values, while pairs sharing only the same strategy but different models span a wider range.
This shows that the solvable frontier is shaped primarily by model identity: switching from LLM-Only to Basic within the same model yields only marginal changes in \emph{which} problems are solved, even when the aggregate count shifts noticeably.
%
\paragraph{Failure mode decomposition.}
We disaggregate failures into timeouts and \texttt{UNKNOWN} outcomes (\texttt{UNKNOWN} means that either ESBMC could neither prove safety nor find a counterexample or another system error occurred either in or outside of ESBMC).
For the two strong models, \texttt{UNKNOWN} rates are low ($\leq 2\%$) and change negligibly under ToT (GPT-OSS-120B: 1.8\% under both Basic and Basic-ToT).
Qwen2.5-7B's \texttt{UNKNOWN} rate under LLM-Only is 26.0\%, but falls to 6.9\% under LLM-Only-ToT ($-19$\,pp); the solve-count nonetheless decreases ($319 \to 316$) as timeouts rise by $+19.5$\,pp.
Llama-3.1-8B shows the reverse: its \texttt{UNKNOWN} rate rises from 4.5\% under Basic to 14.5\% under Basic-ToT (+10\,pp).

These two weak models thus face qualitatively different bottlenecks. For Qwen2.5-7B, ToT's structured output template improves format compliance but exhausts the iteration budget; for Llama-3.1-8B, the structured template itself exceeds the model's reliable output capability. In both cases the binding constraint is not reasoning ability but the interplay between output format and iteration budget.

\paragraph{Model capability vs.\ strategy choice.}

To understand the relative contributions of model identity and strategy selection to solve outcomes, we classify each discriminating problem (one where some (model, strategy) cell succeeds and another fails) by which axis drives most of the variance.
Across the four pre-existing benchmark suites, model-driven variance dominates strategy-driven variance by a factor of \textbf{3--7$\times$}:
on \texttt{L4I-SY}, 31 of 54 discriminating problems are classified model-dominated, while only 4 are strategy-dominated (19 mixed);
on \texttt{C2I}, the ratio is 39:10 (26 mixed).
\texttt{L4I-SVC} is the closest to balanced (20:4 with 32 mixed), making it the most informative benchmark for isolating strategy effects.
The gap between the strongest model (GPT-OSS-120B) and the weakest (Llama-3.1-8B) on their respective best strategies is 89 problems (431 vs.\ 342 under Basic), substantially larger than the within-model strategy span of 67 problems (Basic at 342 vs.\ LLM-Only-ToT at 275 for Llama-3.1-8B).
The within-model maximum gain attributable to strategy alone is $+35$ (Basic vs.\ LLM-Only on Llama-3.1-8B).
These numbers confirm that model capability is the primary driver of accuracy: Phase~1 symbolic feedback provides useful but modest within-model gains, and investing in a stronger model yields a larger expected accuracy gain than any strategy-level refinement.

\paragraph{Compute overhead.}
Table~\ref{tab:tot-comparison} shows that the wall-clock cost of ToT is model-dependent, ranging from negligible to $1.40\times$. For the two strongest models the overhead is small (under 15\%) and the diversity gain is correspondingly modest (1 and 6 ToT-unique pairs); the extra compute is unlikely to be worthwhile in practice unless the specific unsolved problems are known to matter. For Qwen2.5-7B, the picture is reversed: ToT-unique pairs number 24 and the overhead reaches $1.40\times$ under LLM-Only-ToT, yet the aggregate solve count does not improve, consistent with the failure-mode analysis above showing that timeouts absorb the scouting budget. In all cases, the cost of switching from LLM-Only to Basic is negligible: Phase~1 symbolic enumeration together with the atom-level filtering and recombination of each LLM proposal completes in time that is small relative to the LLM refinement loop, so the accuracy gains reported in RQ1 come effectively for free.

%% file: results_table_cot.tex
%
\centering
\begin{tabular}{ccccccccc}
\hline
\multirow{2}{*}{\textbf{LLM Model}} & \multirow{2}{*}{\textbf{Strategy}} & \multicolumn{5}{c}{\textbf{Benchmarks}} & \multirow{2}{*}{\textbf{IR}} & \multirow{2}{*}{\textbf{Total}} \\
\cline{3-7}
 &  & \textbf{C2I} & \textbf{L4I-SVC} & \textbf{L4I-SY} & \textbf{SVC} & \textbf{NL} & & \\

\hline
\multirow{2}{*}{GPT-OSS-120B}
  & Basic    & 127 & 87 & 77 & 85 & 41 & 14 & 431 \\
\cline{2-9}

  & LLM-Only & 127 & 87 & 76 & 82 & 42 & 14 & 428 \\
\hline

\multirow{2}{*}{GPT-OSS-20B}
  & Basic    & 125 & 87 & 73 & 85 & 40 & 14 & 424 \\
\cline{2-9}

  & LLM-Only & 124 & 80 & 71 & 82 & 38 & 14 & 409 \\
\hline

\multirow{2}{*}{Llama-3.1-8B}
  & Basic    & 111 & 71 & 57 & 69 & 22 & 12 & 342 \\
\cline{2-9}

  & LLM-Only &  98 & 65 & 45 & 68 & 18 & 13 & 307 \\
\hline

\multirow{2}{*}{Qwen2.5-32B}
  & Basic    & 112 & 80 & 68 & 76 & 34 & 12 & 382 \\
\cline{2-9}

  & LLM-Only & 113 & 78 & 67 & 75 & 34 & 13 & 380 \\
\hline

\multirow{2}{*}{Qwen2.5-7B}
  & Basic    & 104 & 74 & 62 & 73 & 28 & 11 & 352 \\
\cline{2-9}

  & LLM-Only & 101 & 69 & 52 & 71 & 26 &  9 & 328 \\
\hline
\end{tabular}








%% file: sota_tool_table.tex
%
%
\centering
\resizebox{\linewidth}{!}{%
\begin{tabular}{ccccccccc}
\hline
\multirow{2}{*}{\textbf{Tool}} & \multicolumn{4}{c}{\textbf{Shared benchmarks}} & \multirow{2}{*}{\textbf{Sub\-total$^\ddagger$}} & \multicolumn{2}{c}{\textbf{SV-COMP}} & \multirow{2}{*}{\textbf{Total$^\dagger$}} \\
\cline{2-5}\cline{7-8}
 & \textbf{C2I} & \textbf{L4I-SVC} & \textbf{L4I-SY} & \textbf{NL} & & \textbf{Comp.\,(43)} & \textbf{Full\,(154)} & \\
\hline

ESBMC-Kind & 71 & 14 & 13 & 4 & 102 & 5 & 43 & 145 \\

\hline

2ls & 85 & 27 & 21 & 4 & 137 & 14 & 57 & 194 \\

\hline

CPA-Checker & 97 & 55 & 46 & 8 & 206 & 19 & 76 & 282 \\

\hline

Code2inv & 110 & 47 & 53 & N/A & 210 & 11 & --$^*$ & --$^*$ \\

\hline

Clause2Inv & 132 & 98 & 82 & 44 & \textbf{356} & 30 & --$^*$ & --$^*$ \\

\hline

LaM4Inv & 133 & 95 & 81 & 29 & 338 & 30 & --$^*$ & --$^*$ \\

\hline

LORIS (GPT-4.1) & 131 & 94 & 81 & 45 & 351 & \textbf{39} & --$^*$ & --$^*$ \\

\hline

VerIbmc (best single cell) & 133 & 94 & 78 & 41 & 346 & 36 & \textbf{85} & \textbf{431} \\

\hline

VerIbmc* (4 strats $\times$ 5 models) & 133 & 94 & 79 & 42 & 348 & 36 & \textbf{89} & \textbf{437} \\

\hline
\end{tabular}%
}

%% file: results_table_tot.tex
%
\centering
\begin{tabular}{ccccccccc}
\hline
\multirow{2}{*}{\textbf{LLM Model}} & \multirow{2}{*}{\textbf{Strategy}} & \multicolumn{5}{c}{\textbf{Benchmarks}} & \multirow{2}{*}{\textbf{IR}} & \multirow{2}{*}{\textbf{Total}} \\
\cline{3-7}
 &  & \textbf{C2I} & \textbf{L4I-SVC} & \textbf{L4I-SY} & \textbf{SVC} & \textbf{NL} & & \\
\hline

\multirow{2}{*}{GPT-OSS-120B}
  & Basic-ToT    & 123 & 87 & 75 & 83 & 39 & 14 & 421 \\
\cline{2-9}
  & LLM-Only-ToT & 124 & 87 & 77 & 83 & 38 & 14 & 423 \\
\hline

\multirow{2}{*}{GPT-OSS-20B}
  & Basic-ToT    & 121 & 84 & 73 & 80 & 39 & 14 & 411 \\
\cline{2-9}
  & LLM-Only-ToT & 122 & 86 &  72 & 82 & 38 & 14 &  414 \\
\hline

\multirow{2}{*}{Llama-3.1-8B}
  & Basic-ToT    &  83 & 62 & 49 & 65 & 17 &  9 & 285 \\
\cline{2-9}
  & LLM-Only-ToT &  81 & 56 & 41 & 65 & 21 & 11 & 275 \\
\hline

\multirow{2}{*}{Qwen2.5-32B}
  & Basic-ToT    &  98 & 74 & 64 & 72 & 28 & 13 & 349 \\
\cline{2-9}
  & LLM-Only-ToT & 100 & 69 & 64 & 72 & 30 & 12 & 347 \\
\hline

\multirow{2}{*}{Qwen2.5-7B}
  & Basic-ToT    &  99 & 76 & 58 & 74 & 24 & 10 & 341 \\
\cline{2-9}
  & LLM-Only-ToT &  90 & 72 & 56 & 74 & 24 & 11 & 327 \\
\hline

\end{tabular}








%% file: tot_comparison_table.tex
\centering
\begin{tabular}{lcccccccc}
\hline
\multirow{2}{*}{\textbf{Model}}
  & \multicolumn{3}{c}{\textbf{Basic}}
  & \multicolumn{3}{c}{\textbf{LLM-Only}}
  & \multirow{2}{*}{\shortstack{\textbf{ToT-}\\\textbf{unique}}} \\
\cline{2-7}
 & \textbf{CoT} & \textbf{ToT} & $\boldsymbol{\Delta}$
 & \textbf{CoT} & \textbf{ToT} & $\boldsymbol{\Delta}$ & \\
\hline
GPT-OSS-120B & 431 (5.2\,h) & 421 (5.9\,h) & $-10$ & 428 (4.7\,h) & 423 (5.6\,h) & $-5$  & 1  \\
GPT-OSS-20B  & 424 (5.5\,h) & 411 (6.2\,h) & $-13$ & 409 (6.5\,h) & 414 (6.2\,h) & $+5$  & 6  \\
Llama-3.1-8B & 342 (8.6\,h) & 285 (8.8\,h) & $-57$ & 307 (10.0\,h) & 275 (9.3\,h) & $-32$ & 10 \\
Qwen2.5-32B  & 382 (6.3\,h) & 349 (7.0\,h) & $-33$ & 380 (6.2\,h) & 347 (6.8\,h) & $-33$ & 8  \\
Qwen2.5-7B   & 352 (6.6\,h) & 341 (7.6\,h) & $-11$ & 328 (5.7\,h) & 327 (8.0\,h) & $-1$  & 24 \\
\hline
\end{tabular}

%% file: appendix/appendix.tex
\section{SV-COMP Frontend Limitations}
\label{sec:svc-limitation-appendix}

Table~\ref{tab:svc-limitation} details why 71\% of the SV-COMP loop set (110 of 154 programs) cannot be expressed by the \texttt{clang-fe} frontend used by Clause2Inv, LaM4Inv, and LORIS. Each cause is an architectural boundary of the frontend or single-loop scalar program graph model, not a tuning gap.

\begin{table}[h]
\centering
\input{svc_limitation_table}
\caption{Breakdown of the 110 SV-COMP loop programs inexpressible by the \texttt{clang-fe} frontend (used by Clause2Inv, LaM4Inv, and LORIS). Each cause is an architectural boundary, not a tuning gap.}
\label{tab:svc-limitation}
\end{table}

\newpage
\section{Additions to Discussion of Experimental Results}

\paragraph{Phase-level ablation.}
Table~\ref{tab:ablation} shows the per-phase breakdown of solve counts for each model under the LLM-Only and Basic strategies.

\begin{table}[h]
\centering
\begin{tabular}{llrrrr}
\hline
\textbf{Model} & \textbf{Strategy} & \textbf{Phase~0} & \textbf{Phase~1} & \textbf{Phase~2} & \textbf{Total} \\
\hline
\multirow{2}{*}{GPT-OSS-120B}
 & LLM-Only  & 111 &   - & 303 & 414 \\
 & Basic     & 111 &  75 & 228 & 414 \quad ($\Delta_{\text{Ph1}}=0$) \\
\hline
\multirow{2}{*}{GPT-OSS-20B}
 & LLM-Only  & 111 &   - & 284 & 395 \\
 & Basic     & 111 &  75 & 224 & 410 \quad ($\Delta_{\text{Ph1}}=+15$) \\
\hline
\multirow{2}{*}{Qwen2.5-32B}
 & LLM-Only  & 111 &   - & 256 & 367 \\
 & Basic     & 111 &  75 & 184 & 370 \quad ($\Delta_{\text{Ph1}}=+3$) \\
\hline
\multirow{2}{*}{Qwen2.5-7B}
 & LLM-Only  & 111 &   - & 208 & 319 \\
 & Basic     & 111 &  75 & 155 & 341 \quad ($\Delta_{\text{Ph1}}=+22$) \\
\hline
\multirow{2}{*}{Llama-3.1-8B}
 & LLM-Only  & 111 &   -  & 183 & 294 \\
 & Basic     & 111 &  75 & 144 & 330 \quad ($\Delta_{\text{Ph1}}=+36$) \\
\hline
\end{tabular}
\caption{Phase-level ablation of \ourtool{} (single-LLM strategies).
\emph{Phase~0}: problems solved by ESBMC without any invariant (standard semantics, model-invariant).
\emph{Phase~1}: problems solved by symbolic candidate conjunction alone (no LLM call, model-invariant at 75 problems).
\emph{Phase~2}: problems requiring at least one LLM-generated invariant.
$\Delta_{\text{Ph1}}$ = net gain of adding Phase~1 (Basic vs.\ LLM-Only).
Dual-LLM data is in the remainder of this appendix.}
\label{tab:ablation}
\end{table}

Figure~\ref{fig:itercdf} reports number of iterations required.
The long tail past 20 refinement rounds is universally rare ($<7\%$ of LLM-solved instances), which provides an empirical basis for capping \texttt{max\_iter} at a modest value (e.g., $\leq 20$).
Without ESBMC Phase~1 feedback (LLM-Only), weaker models continue to gain solves well past cap~$=20$, confirming that the symbolic augmentation in Basic narrows the long tail even without a second LLM.

\begin{table}
\centering
\begin{tabular}{llrrrrrrr}
\hline
\textbf{Model} & \textbf{Strategy} & \textbf{N} & \textbf{$\leq 0$} & \textbf{$\leq 1$} & \textbf{$\leq 5$} & \textbf{$\leq 10$} & \textbf{$\leq 20$} & \textbf{p90} \\
\hline
\multirow{4}{*}{GPT-OSS-120B} & Basic & 236 & 83.1 & 91.1 & 98.7 & 99.6 & 100.0 & 1 \\
 & Basic-ToT & 228 & 69.7 & 74.6 & 92.5 & 99.6 & 100.0 & 5 \\
 & LLM-Only & 268 & 72.0 & 91.8 & 99.3 & 100.0 & 100.0 & 1 \\
 & LLM-Only-ToT & 310 & 78.1 & 83.5 & 94.5 & 100.0 & 100.0 & 5 \\
\hline
\multirow{4}{*}{GPT-OSS-20B} & Basic & 231 & 77.5 & 89.2 & 97.4 & 99.6 & 100.0 & 1 \\
 & Basic-ToT & 218 & 77.1 & 78.9 & 94.5 & 99.5 & 100.0 & 5 \\
 & LLM-Only & 295 & 72.2 & 88.5 & 100.0 & 100.0 & 100.0 & 1 \\
 & LLM-Only-ToT & 301 & 81.4 & 82.7 & 94.0 & 100.0 & 100.0 & 5 \\
\hline
\multirow{4}{*}{Llama-3.1-8B} & Basic & 150 & 26.0 & 36.0 & 50.0 & 62.0 & 73.3 & 36 \\
 & Basic-ToT & 93 & 34.4 & 37.6 & 66.7 & 89.2 & 97.8 & 11 \\
 & LLM-Only & 194 & 24.7 & 35.1 & 50.5 & 60.8 & 69.6 & 38 \\
 & LLM-Only-ToT & 162 & 37.7 & 43.2 & 72.8 & 91.4 & 96.3 & 10 \\
\hline
\multirow{4}{*}{Qwen2.5-32B} & Basic & 189 & 57.7 & 67.7 & 76.2 & 80.4 & 86.8 & 33 \\
 & Basic-ToT & 156 & 57.7 & 62.8 & 84.6 & 91.0 & 94.9 & 9 \\
 & LLM-Only & 266 & 49.6 & 59.0 & 70.3 & 74.8 & 81.2 & 40 \\
 & LLM-Only-ToT & 233 & 62.2 & 65.2 & 84.5 & 96.6 & 98.3 & 9 \\
\hline
\multirow{4}{*}{Qwen2.5-7B} & Basic & 159 & 35.8 & 43.4 & 54.1 & 59.7 & 78.0 & 17 \\
 & Basic-ToT & 148 & 41.9 & 45.3 & 64.9 & 77.7 & 83.8 & 33 \\
 & LLM-Only & 214 & 40.2 & 47.7 & 54.7 & 58.9 & 72.4 & 27 \\
 & LLM-Only-ToT & 213 & 50.7 & 52.6 & 70.9 & 85.0 & 90.6 & 17 \\
\hline
\end{tabular}
\caption{Per-cell convergence thresholds. \textbf{N}: number of successfully-solved problems where at least one LLM call was made; excludes Phase~0-only and Phase~1-only solves, but includes integer-relaxed solves. \textbf{$\leq k$}: cumulative percentage of those $N$ problems solved within $k$ LLM refinement rounds ($k{=}0$ means first LLM call succeeded). \textbf{p90}: 90th-percentile round count. Convergence data for the dual-LLM strategies is reported in \prettyref{sec:dual-llm-supp}.}
\label{tab:iter-cdf}
\end{table}

\section{Coverage}
\begin{table}
\centering
\begin{tabular}{lrrr}
\hline
\textbf{Strategy} & \textbf{Solved} & \textbf{Coverage \%} & \textbf{Unique vs others} \\
\hline
Basic & 317 & 97.8 & 1 \\
Basic-ToT & 315 & 97.2 & 1 \\
LLM-Only & 314 & 96.9 & 0 \\
LLM-Only-ToT & 318 & 98.1 & 1 \\
\hline
\textbf{Union (any strategy)} & \textbf{324} & \textbf{100.0} & \textbf{--} \\
\hline
\end{tabular}
\caption{Strategy-as-solver coverage of the 323 non-trivial problems reachable by at least one of the four inference strategies (Phase~0-safe and dual-LLM-exclusive problems excluded). Each strategy = union of solves across all five models. ``Unique vs others'' counts problems that no other single-LLM strategy (in its own union) reaches.}
\label{tab:strategy-frontier}
\end{table}

\newpage
\section{Dual-LLM Correction with a Fixed Strong Model}
\label{sec:dual-llm-supp}

\paragraph{Strategy description.}
The Combined strategy follows the same Phase~0--Phase~1 structure as the Basic strategy but replaces the single LLM in Phase~2 with a \emph{dual LLM sequence}.
In each iteration, LLM$_1$ receives the standard prompt and generates invariant candidates; its raw textual output together with the current $\mathcal{P}$, $\mathcal{D}$, $\mathcal{U}$ lists is forwarded as context to LLM$_2$.
LLM$_2$ can therefore observe both what LLM$_1$ proposed and what ESBMC formally verified, enabling semantic-level correction rather than blind regeneration.
Both outputs are independently parsed, verified, and atomically decomposed; a successful outcome may arise from LLM$_1$ alone, LLM$_2$ alone, or the recombination of atoms from both.
The Combined-ToT variant is \emph{not} a simple extension of Combined; it replaces the sequential per-iteration correction chain with the Tree-of-Thoughts multi-branch scouting structure (\S3.7).
In each ToT stage, both LLM$_1$ and LLM$_2$ are invoked across multiple exploratory branches, and their accumulated atoms jointly populate the shared stores $\mathcal{P}$, $\mathcal{D}$, $\mathcal{U}$; a top-2 selection then prunes the branch set before the next stage.
The result is that LLM$_2$ no longer acts as a targeted corrector of LLM$_1$'s output within a single iteration---instead, both models contribute to cross-branch diversity, at the cost of a larger iteration budget.
In short, the two strategies differ not merely in the presence of ToT but in how the two LLMs interact: LLM$_2$ is a sequential corrector in Combined and a joint explorer in Combined-ToT.

\paragraph{Setup.}

Important: the Combined and Combined-ToT strategies each use two LLMs simultaneously and are not directly comparable to the four single-LLM strategies above.
The gains below scale with the capability gap between LLM$_1$ and GPT-OSS-120B; they should be read as a characterisation of strong-model post-processing as an optional ceiling, not as evidence that dual-LLM refinement is a generally effective technique.
LLM$_2$ is fixed to GPT-OSS-120B---the strongest model in our evaluation pool---regardless of which model serves as LLM$_1$.
We report these results separately as a supplement.

\paragraph{Aggregate gains.}

\begin{table}
\centering
\begin{tabular}{lrrr}
\hline
\textbf{LLM$_1$} & \textbf{Basic} & \textbf{Combined} & \textbf{$\Delta$} \\
\hline
GPT-OSS-120B (= LLM$_2$) & 428 & 427 & $-1$ \\
GPT-OSS-20B               & 424 & 425 & $+1$ \\
Llama-3.1-8B               & 342 & \textbf{389} & $+47$ \\
Qwen2.5-32B               & 382 & \textbf{421} & $+39$ \\
Qwen2.5-7B                & 352 & \textbf{416} & $+64$ \\
\hline
\end{tabular}
\caption{Supplementary dual-LLM experiment: Basic vs.\ Combined solve counts (out of 499 effective problems). LLM$_2$ = GPT-OSS-120B is fixed for all runs. $\Delta$ = combined $-$ basic.}
\label{tab:dual-llm-gains}
\end{table}

Table~\ref{tab:dual-llm-gains} shows the gains of Combined over Basic for each LLM$_1$.
When LLM$_1$ = LLM$_2$ = GPT-OSS-120B, the correction adds zero benefit ($\Delta = -1$); as LLM$_1$ weakens, the gain grows monotonically to $+64$ for Qwen2.5-7B.
This pattern indicates that the gain tracks the capability gap between LLM$_1$ and GPT-OSS-120B rather than reflecting any benefit of ESBMC feedback structure, which explains why the two experimental conditions cannot be compared on equal terms.

\paragraph{Unique solves.}

\begin{figure}[h]
\centering
\includegraphics[width=0.85\columnwidth]{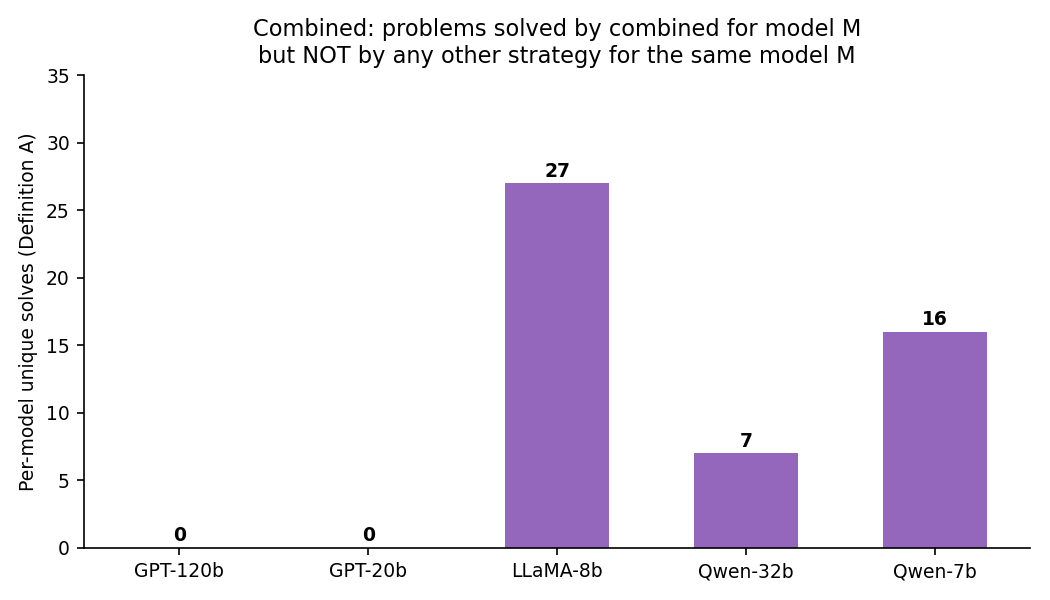}
\caption{Per-model unique solves (Definition A): problems solved by Combined for model $M$ but not by any other strategy for the same model $M$. Total = 50 (model, problem) pairs.}
\label{fig:unique-combined}
\end{figure}

\begin{figure}[h]
\centering
\includegraphics[width=0.95\columnwidth]{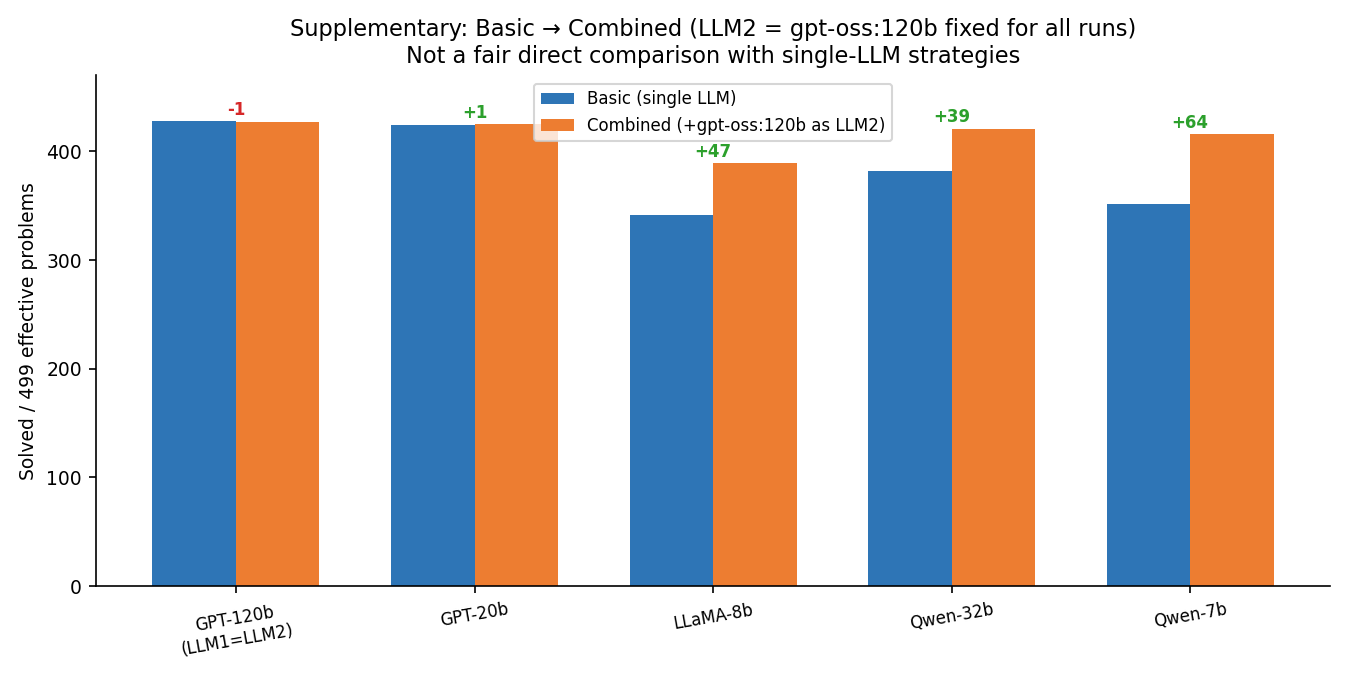}
\caption{Supplementary dual-LLM experiment: Basic (single LLM) vs.\ Combined (+GPT-OSS-120B as LLM$_2$). Numbers show the $\Delta$ (combined $-$ basic). Not directly comparable to the four single-LLM strategies.}
\label{fig:dual-llm-supp}
\end{figure}

Despite the aggregate gain being largely attributable to LLM$_2$'s capability, Combined's solution trajectory---which begins from LLM$_1$'s partial invariant proposals---reaches problems that no other strategy for the same model can solve.
Under Definition A (per-model unique: solved by Combined for model $M$ but by no other strategy for $M$), the counts are 27 for Llama-3.1-8B, 16 for Qwen2.5-7B, and 7 for Qwen2.5-32B (50 model--problem pairs in total; Figure~\ref{fig:unique-combined}).

Under the stricter Definition B---where no other (model, strategy) pair can solve the problem---only 2 problems are globally unique to Combined, both solved by Qwen2.5-7B: \texttt{lam4inv\_sygus19/186.c} and \texttt{sv-comp\_loops-crafted-1/mono-crafted\_3.c}. These two problems account for the difference between the union of 437 (four strategies) and 440 (all six configurations).

\paragraph{Convergence.}

Under Combined, first-try success rates are uniformly high ($\geq 77\%$ across all models), compared to 24--83\% under Basic.
This faster convergence is explained by LLM$_2$ (GPT-OSS-120B) frequently reformatting or correcting LLM$_1$'s output on the first pass, reducing the number of ESBMC re-queries required.
Consequently, the low UNKNOWN (parse failure) rate observed for weaker models under Combined reflects GPT-OSS-120B's output quality rather than any structural benefit of ESBMC feedback.


\section{Similar-Size Paired LLM Experiments}
\label{sec:similar-size-supp}

\paragraph{Motivation.}
The experiment in \prettyref{sec:dual-llm-supp} holds LLM$_2$ fixed to GPT-OSS-120B, the strongest model in our pool, so the observed gains track the \emph{capability gap} between LLM$_1$ and GPT-OSS-120B rather than a structural benefit of dual-LLM correction.
To isolate the structural question, we re-run the Combined and Combined-ToT strategies with four LLM pairs whose two members are drawn from the \emph{same capability tier}: two small-model pairs ($\approx$7--8B parameters) and two medium-model pairs ($\approx$20--32B parameters).
LLM$_2$ is \textbf{not} GPT-OSS-120B in any of these runs; the same 520-problem pool and 600\,s timeout are used throughout.

\paragraph{Important.}
\textbf{Combined and Combined-ToT use architecturally different pipelines} (see \S\ref{sec:dual-llm-supp}, Strategy distinction) and are not directly comparable to each other or to the four single-LLM strategies.
Results are reported to characterise dual-LLM behaviour in the equal-size regime, not as a controlled ablation.

\begin{table*}[t]
\centering
\caption{Similar-size paired LLM results across all 11 benchmarks (520 problems).
\textbf{C} = Combined, \textbf{CT} = Combined-ToT.
IR-mode solves are noted separately in the Total row (not added to per-benchmark columns).
Contrast with Table~\ref{tab:dual-llm-gains}, where LLM$_2$ = GPT-OSS-120B throughout.
Pair labels: \textbf{P1} = Qwen-7B$\to$LLaMA-8B; \textbf{P2} = LLaMA-8B$\to$Qwen-7B; \textbf{P3} = Qwen-32B$\to$GPT-20B; \textbf{P4} = GPT-20B$\to$Qwen-32B.}
\label{tab:similar-size}
\resizebox{\linewidth}{!}{%
\begin{tabular}{l rr rr rr rr}
\toprule
 & \multicolumn{4}{c}{\textbf{Small ($\approx$7--8B)}} & \multicolumn{4}{c}{\textbf{Medium ($\approx$20--32B)}} \\
\cmidrule(lr){2-5}\cmidrule(lr){6-9}
 & \multicolumn{2}{c}{\textbf{P1}} & \multicolumn{2}{c}{\textbf{P2}} & \multicolumn{2}{c}{\textbf{P3}} & \multicolumn{2}{c}{\textbf{P4}} \\
\cmidrule(lr){2-3}\cmidrule(lr){4-5}\cmidrule(lr){6-7}\cmidrule(lr){8-9}
\textbf{Benchmark (total)} & \textbf{C} & \textbf{CT} & \textbf{C} & \textbf{CT} & \textbf{C} & \textbf{CT} & \textbf{C} & \textbf{CT} \\
\midrule
loop-acceleration (24) & 17 & 14 & 15 & 15 & 20 & 15 & 19 & 17 \\
loop-crafted (8)        &  2 &  2 &  2 &  2 &  2 &  2 &  2 &  2 \\
loop-invariants (9)     &  6 &  3 &  6 &  4 &  6 &  6 &  6 &  6 \\
loop-new (11)           &  2 &  2 &  2 &  2 &  2 &  2 &  2 &  2 \\
loop-simple (6)         &  5 &  5 &  5 &  5 &  5 &  5 &  5 &  5 \\
loops (50)              & 25 & 23 & 23 & 24 & 26 & 25 & 25 & 25 \\
loops-crafted-1 (46)    & 15 & 16 & 16 & 13 & 23 & 20 & 24 & 19 \\
\midrule
C2I (133)    & 91 & 84 & 84 & 87 & \textbf{133} & 126 & 132 & 121 \\
L4I SVC (99)  & 78 & 71 & 67 & 59 & 93 & 85 & 93 & 87 \\
L4I SY (84)   & 57 & 44 & 53 & 35 & 78 & 76 & 76 & 77 \\
NL (50)     & 22 & 15 & 19 & 17 & 38 & 33 & 39 & 38 \\
\midrule
\textbf{Total / 520}    & \textbf{320} & \textbf{279} & \textbf{292} & \textbf{263} & \textbf{426} & \textbf{395} & \textbf{423} & \textbf{399} \\
\quad\textit{of which IR} & \textit{(+10)} & \textit{(+9)} & \textit{(+9)} & \textit{(+9)} & \textit{(+14)} & \textit{(+13)} & \textit{(+14)} & \textit{(+11)} \\
\textbf{Solve \%}       & 61.5 & 53.7 & 56.2 & 50.6 & 81.9 & 76.0 & 81.3 & 76.7 \\
\bottomrule
\end{tabular}}
\end{table*}

\paragraph{Findings.}
Three patterns emerge from Table~\ref{tab:similar-size}.
\textbf{(i) Model size dominates.}
Medium pairs solve 81--82\% of problems versus 56--62\% for small pairs---a gap of $\approx$130 problems---mirroring the main-paper conclusion that model capability is the primary accuracy driver even in the equal-size regime.
\textbf{(ii) LLM1/LLM2 order sensitivity is weak.}
Swapping the two models changes the aggregate by at most 28 problems for small pairs (P1 Combined: 320 vs.\ P2 Combined: 292) and at most 3 problems for medium pairs, indicating that the correction gain depends on the shared capability tier rather than on which model sees the problem first.
\textbf{(iii) Combined-ToT consistently underperforms Combined} across all four pairs (by 32--54 problems for small pairs, 24--31 for medium pairs), consistent with the budget-cost analysis in \prettyref{sec:dual-llm-supp}: when neither LLM is large enough to absorb the iteration overhead of multi-branch scouting, ToT's diversity benefit does not materialise.
Notably, Pair~3 (Qwen2.5-32B + GPT-OSS-20B) achieves 133/133 (100\%) on \texttt{ibmc\_code2inv} under Combined, matching the ceiling attained by GPT-OSS-120B alone in the main experiment.

%
%
%
%
%
%

%% file: svc_limitation_table.tex
\centering
\begin{tabular}{lrl}
\hline
\textbf{Reason cannot be expressed} & \textbf{Count} & \textbf{Inherent cause} \\
\hline
Array              & 41 & graph models only scalar integers \\
Nested/multi-loop  & 36 & approach targets single loops \\
Pointer            &  5 & pointers not modeled \\
Frontend crash     &  8 & \texttt{clang-fe} hang/segfault \\
In-loop assert     &  8 & empty post-condition; Horn template cannot encode \\
Other VC bug       & 11 & malformed/ill-typed VC; function calls, infinite loops \\
Bitwise            &  1 & not expressible in linear integer arithmetic \\
\hline
\textbf{Total limitation} & \textbf{110} & (71\% of 154) \\
\hline
\end{tabular}